\documentclass[prx,aps,letterpaper,twocolumn,allowfontchageintitle=true,accepted=2019-10-30]{quantumarticle}

\usepackage{bbm}
\usepackage{overpic}
\usepackage{amsmath}
\usepackage{epsfig,psfrag}
\usepackage{multirow}
\usepackage{dcolumn}% Align table columns on decimal point
\usepackage{bm}% bold math
\usepackage{graphicx}
\usepackage{color}
\usepackage{version}
\usepackage[allcolors=quantumviolet,pdftitle={Magic State Distillation: Not as Costly as You Think},pdfauthor={Daniel Litinski}]{hyperref}
\usepackage{natbib}
\usepackage{multicol}

\usepackage{float}
\usepackage{wasysym}
\usepackage{algorithm2e}
\SetAlgoCaptionSeparator{}
\SetAlCapSkip{0.5em}
\NoCaptionOfAlgo

\usepackage{xcolor}
\usepackage{braket}
\begin{document}

\title{\centering \color{quantumviolet} Magic State Distillation: Not as Costly as You Think \newpage}
\author{\vspace{-6ex}}
\affiliation{Daniel Litinski @ Dahlem Center for Complex Quantum Systems, Freie Universit\"at Berlin, Arnimallee 14, 14195 Berlin, Germany}

\date{\vspace{-5ex}}

{\centering \maketitle}

\begin{abstract}

Despite significant overhead reductions since its first proposal, magic state distillation is often considered to be a very costly procedure that dominates the resource cost of fault-tolerant quantum computers. The goal of this work is to demonstrate that this is not true. By writing distillation circuits in a form that separates qubits that are capable of error detection from those that are not, most logical qubits used for distillation can be encoded at a very low code distance. This significantly reduces the space-time cost of distillation, as well as the number of qubits. In extreme cases, it can cost less to distill a magic state than to perform a logical Clifford gate on full-distance logical qubits.

\end{abstract}

\begin{figure}[!b]
\centering
\def\svgwidth{0.9\linewidth}
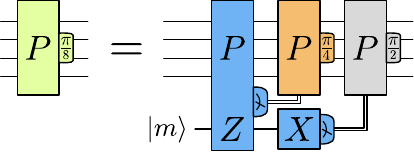
\caption{A $P_{\pi/8}$ rotation on $n$ qubits can be performed by measuring the Pauli product operator $P \otimes Z$ acting on the $n$ qubits and a magic state $\ket{m} = (\ket{0} + e^{i\pi/4}\ket{1})/\sqrt{2}$. The magic state is discarded via an $X$ measurement. Measurement outcomes of $-1$ of the $P \otimes Z$ or $X$ measurement prompt a $P_{\pi/4}$ or $P_{\pi/2}$ correction, respectively.}
\label{fig:magicstateuse}
\end{figure}

Quantum error correction is expected to be an essential part of a large-scale quantum computer~\cite{Preskill1998,TerhalRMP,Campbell2016}. The most promising quantum error-correcting codes are two-dimensional topological codes such as surface codes~\cite{Kitaev2003,Fowler2012}. With these codes, the preparation of logical Pauli eigenstates and the measurement of logical Pauli product operators (e.g., via lattice surgery~\cite{Horsman2012,Litinski2017b,Fowler2018}) are fault-tolerant operations, which is sufficient for fault-tolerant logical Clifford gates. However, logical non-Clifford gates, such as $T$ gates, cannot be executed directly. Instead, they can be performed by preparing a resource state that is consumed to execute a non-Clifford gate~\cite{Bravyi2005}. For $T$ gates, this resource state is a magic state $\ket{m} = (\ket{0} + e^{i\pi/4}\ket{1})/\sqrt{2}$. Such a state can be used to perform a $\pi/8$ rotation $P_{\pi/8} = e^{-iP\pi/8}$, where $P$ is an $n$-qubit Pauli product operator from the set $\{X,Y,Z,\mathbbm{1}\}^{\otimes n}$. Here, $X$, $Y$ and $Z$ are Pauli operators, such that $Z_{\pi/8}$ corresponds to a $T$ gate. If a magic state is available, a logical $n$-qubit $P_{\pi/8}$ gate is performed by measuring the logical Pauli product $P \otimes Z$ acting on the $n$ qubits and the magic state, see Fig.~\ref{fig:magicstateuse}.

The problem with magic states is that, with surface codes, only faulty magic states can be prepared. They are faulty in the sense that they are initialized with an error probability proportional to the physical error rate $p_{\rm phys}$, regardless of the code distance. If these states are used to perform logical $P_{\pi/8}$ rotations, one out of every ${\sim}1/p_{\rm phys}$ logical gates is expected to be faulty. Since faulty gates spoil the outcome of a computation, but classically intractable quantum computations with a useful computational result typically involve more than $10^8$ $T$ gates~\cite{Babbush2018}, low-error magic states are required to execute gates with a low error probability. One possibility to generate low-error magic states is via a magic state distillation protocol. These protocols are short error-detecting quantum computations that use multiple high-error magic states to generate fewer low-error states. Many such protocols~\cite{Bravyi2012,Fowler2013,Meier2013,Jones2013a,DuclosCianci2013,Duclos2015,Campbell2017,Ogorman2017,Haah2018,Campbell2018,Fowler2018,Gidney2018a} have been developed since magic state distillation was first proposed~\cite{Bravyi2005}, gradually decreasing the cost of distillation. Even though state-of-the-art protocols are orders of magnitude more efficient than the earliest proposals, magic state distillation is still often described as a costly procedure and the leading contributing factor to the overhead of fault-tolerant quantum computing, which is the primary motivation for research into alternatives to magic state distillation~\cite{Jones2016,Bravyi2015,Jochym2016,Bombin2018,Chamberland2018,Brown2019}.

In this work, we reduce the cost of distillation by another order of magnitude. On the level of circuits, none of the distillation protocols discussed in this work are new. Rather, the circuits are written in a way that the number of qubits is low and the circuit depth is high. The overhead reduction is achieved by finding surface-code implementations of these protocols in which the code distance of each surface-code patch is never higher than required to achieve a specific output error probability, as was previously proposed in Ref.~\cite{Gidney2018a}. This yields protocols that not only have a low space-time cost, but also a small qubit footprint.

\renewcommand{\arraystretch}{1.2}

\begin{table*}[t]
\centering
\scalebox{0.91}{
\begin{tabular}{c|c|l|c|c|c|c}

\multirow{2}{*}{Protocol} & \multirow{2}{*}{$p_{\rm phys}$} & \multicolumn{1}{c|}{\multirow{2}{*}{$p_{\rm out}$}} & \multirow{2}{*}{Qubits} & \multirow{2}{*}{Cycles} & \multicolumn{2}{c}{Space-time cost per output state} \\
& & & & & Qubitcycles & Full distance \\
\cline{1-7}
$\text{(15-to-1)}_{7,3,3}$ & $10^{-4}$ & $4.4 \times 10^{-8}$ & 810 & 18.1 & 14,600 &
$5.49d^3 \hspace{0.1em} / \hspace{0.1em} d=11$ \vline~$3.33d^3 \hspace{0.1em} / \hspace{0.1em} d=13$  \\ 

$\text{(15-to-1)}_{9,3,3}$ & $10^{-4}$ & $9.3 \times 10^{-10}$ & 1,150 & 18.1 & 20,700 & 
$4.71d^3 \hspace{0.1em} / \hspace{0.1em} d=13$ \vline~$3.07d^3 \hspace{0.1em} / \hspace{0.1em} d=15$   \\

$\text{(15-to-1)}_{11,5,5}$ & $10^{-4}$ & $1.9 \times 10^{-11}$ & 2,070 & 30.0 & 62,000 & 
$9.19d^3 \hspace{0.1em} / \hspace{0.1em} d=15$ \vline~$6.31d^3 \hspace{0.1em} / \hspace{0.1em} d=17$  \\

$\text{(15-to-1)}_{9,3,3}^4 \times \text{(20-to-4)}_{15,7,9}$ & $10^{-4}$ & $2.4 \times 10^{-15}$ & 16,400 & 90.3 & 371,000 & $27.0d^3 \hspace{0.1em} / \hspace{0.1em} d=19$ \vline~$20.0d^3 \hspace{0.1em} / \hspace{0.1em} d=21$  \\

$\text{(15-to-1)}_{9,3,3}^4 \times \text{(15-to-1)}_{25,9,9}$ & $10^{-4}$ & $6.3 \times 10^{-25}$ & 18,600 & 67.8 & 1,260,000 & 
$25.9d^3 \hspace{0.1em} / \hspace{0.1em} d=29$ \vline~$21.2d^3 \hspace{0.1em} / \hspace{0.1em} d=31$ \\

\cline{1-7}

$\text{(15-to-1)}_{17,7,7}$ & $10^{-3}$ & $4.5 \times 10^{-8}$ & 4,620 & 42.6 & 197,000 & $6.30d^3 \hspace{0.1em} / \hspace{0.1em} d=25$ \vline~$4.04d^3 \hspace{0.1em} / \hspace{0.1em} d=29$ \\

$\text{(15-to-1)}_{13,5,5}^6 \times \text{(20-to-4)}_{23,11,13}$ & $10^{-3}$ & $1.4 \times 10^{-10}$ & 43,300 & 130 & 1,410,000 & $28.9d^3 \hspace{0.1em} / \hspace{0.1em} d=29$ \vline~$19.6d^3 \hspace{0.1em} / \hspace{0.1em} d=33$ \\

$\text{(15-to-1)}_{13,5,5}^4 \times \text{(20-to-4)}_{27,13,15}$ & $10^{-3}$ & $2.6 \times 10^{-11}$ & 46,800 & 157 & 1,840,000 & $30.9d^3 \hspace{0.1em} / \hspace{0.1em} d=31$ \vline~$21.5d^3 \hspace{0.1em} / \hspace{0.1em} d=35$ \\

$\text{(15-to-1)}_{11,5,5}^6 \times \text{(15-to-1)}_{25,11,11}$ & $10^{-3}$ & $2.7 \times 10^{-12}$ & 30,700 & 82.5 & 2,540,000 & $35.3d^3 \hspace{0.1em} / \hspace{0.1em} d=33$ \vline~$25.0d^3 \hspace{0.1em} / \hspace{0.1em} d=37$ \\

$\text{(15-to-1)}_{13,5,5}^6 \times \text{(15-to-1)}_{29,11,13}$ & $10^{-3}$ & $3.3 \times 10^{-14}$ & 39,100 & 97.5 & 3,810,000 & $37.6d^3 \hspace{0.1em} / \hspace{0.1em} d=37$ \vline~$27.7d^3 \hspace{0.1em} / \hspace{0.1em} d=41$ \\

$\text{(15-to-1)}_{17,7,7}^6 \times \text{(15-to-1)}_{41,17,17}$ & $10^{-3}$ & $4.5 \times 10^{-20}$ & 73,400 & 128 & 9,370,000 & $39.8d^3 \hspace{0.1em} / \hspace{0.1em} d=49$ \vline~$31.5d^3 \hspace{0.1em} / \hspace{0.1em} d=53$ \\

\cline{1-7}

\multicolumn{7}{c}{Small-footprint and synthillation protocols} \\

\cline{1-7}

$\text{(15-to-1)}_{9,3,3}$ & $10^{-4}$ & $1.5 \times 10^{-9}$ & 762 & 36.2 & 27,600 &
$6.27d^3 \hspace{0.1em} / \hspace{0.1em} d=13$ \vline~$4.08d^3 \hspace{0.1em} / \hspace{0.1em} d=15$\\

$\text{(15-to-1)}_{9,5,5} \times \text{(15-to-1)}_{21,9,11}$ & $10^{-3}$ & $6.1 \times 10^{-10}$ & 7,780 & 469 & 3,650,000 & $74.7d^3 \hspace{0.1em} / \hspace{0.1em} d=29$ \vline~$50.7d^3 \hspace{0.1em} / \hspace{0.1em} d=33$ \\

$\text{(15-to-1)}_{7,3,3}^4 \times \text{(8-to-CCZ)}_{15,7,9}$ & $10^{-4}$ & $7.2 \times 10^{-14}$ & 12,400 & 36.1 & 447,000 & $32.6d^3 \hspace{0.1em} / \hspace{0.1em} d=19$ \vline~$24.1d^3 \hspace{0.1em} / \hspace{0.1em} d=21$ \\

$\text{(15-to-1)}_{13,7,7}^6 \times \text{(8-to-CCZ)}_{25,15,15}$ & $10^{-3}$ & $5.2 \times 10^{-11}$ & 47,000 & 60.0 & 2,820,000 & $47.4d^3 \hspace{0.1em} / \hspace{0.1em} d=31$ \vline~$32.9d^3 \hspace{0.1em} / \hspace{0.1em} d=35$ \\

\cline{1-7}

\multicolumn{7}{c}{Historical numbers} \\

\cline{1-7}

$\text{(15-to-1)}$ in Ref.~\cite{Litinski2019} & $10^{-4}$ & $3.5 \times 10^{-11}$ & 3,720 & 143 & 532,000 & $121d^3~(d=13)$ \\

$\text{(15-to-1)} \times \text{(8-to-2)}$ in Ref.~\cite{Gidney2018a} & $10^{-3}$ & $2.7 \times 10^{-11}$ & 148,000 & 202 & 14,900,000 & $251d^3~(d=31)$ \\

$\text{(15-to-1)} \times \text{(8-to-CCZ)}$ in Ref.~\cite{Gidney2018a} & $10^{-3}$ & $5.3 \times 10^{-11}$ & 134,000 & 171 & 22,800,000 & $473d^3~(d=31)$ \\

$\text{(15-to-1)} \times \text{(15-to-1)}$ in Ref.~\cite{Fowler2018} & $10^{-3}$ & $1.0 \times 10^{-14}$ & 177,000 & 202 & 35,700,000 & $599d^3~(d=31)$ \\

$\text{(15-to-1)} \times \text{(15-to-1)}$ in Ref.~\cite{Fowler2012} & $10^{-3}$ & $3.0 \times 10^{-15}$ & 800,000 & 250 &  200,000,000  & $2544d^3~(d=34)$ \hspace{0.1em}

\end{tabular}}

\caption{Comparison of different distillation protocols with respect to the following characteristics: physical error rate $p_{\rm phys}$, output error probability per output state $p_{\rm out}$, space cost in qubits, time cost in surface-code cycles, and space-time cost in qubitcycles. The last two columns report the space-time cost in $\text{(physical data qubits)} \times \text{(code cycles)}$ measured in units of the full distance $d^3$, where $d$ is the distance required for the data qubits of a 100-qubit (left column) or 10,000-qubit (right column) computation with at most $1/p_{\rm out}$ $T$ gates. The subscripts and superscripts in the protocol description indicate the code distances and number of level-1 distillation blocks used in the protocol, as explained in Secs.~\ref{sec:15to1}-\ref{sec:smallfootprint}.}
\label{tab:results}
\end{table*}

\textbf{The cost of distillation.} How does one quantify the cost of a distillation protocol? The space-time cost is often quantified in units of $d^3$, where $d$ is the code distance. However, this can be confusing, if different code distances are used in different parts of the quantum computer. Consider an $n_q$-qubit quantum computation with $n_T$ $T$ gates and a quantum computer consisting of a block of qubits used to distill magic states and a block of qubits used to store the $n_q$ data qubits and consume magic states~\cite{Litinski2019}. The distance required for the storage of the data qubits depends on $n_q$ and $n_T$, as it needs to be high enough to guarantee that the probability of an error on any of the $n_q$ qubits during the entire $n_T$-$T$ gate computation is sufficiently low. In other words, this distance is governed by the space-time volume of the computation ${\sim}n_q\cdot n_T$, as weight-$d/2$ error strings in this space-time volume can potentially corrupt the output of the computation. We will refer to this distance as the \textit{full distance}. The code distances used for distillation, on the other hand, are completely irrelevant. The protocol merely needs to produce magic states with an output error probability that is lower than $1/n_T$, for which it uses a certain number of qubits for a certain number of code cycles. Since the full distance depends on $n_q$, but the space-time cost of a distillation protocol does not, it is more meaningful to quantify the space-time cost in terms of qubitcycles, i.e., $\text{qubits} \cdot \text{cycles}$.

\textbf{Results.} Table~\ref{tab:results} shows the space-time costs of the protocols that are constructed in the following sections. These protocols generate states with different output error probabilities $p_{\rm out}$, assuming physical circuit-level error rates $p_{\rm phys}$ of $10^{-3}$ and $10^{-4}$. The more $T$ gates need to be executed, the lower $p_{\rm out}$ needs to be. Each protocol is characterized by the space cost in terms of physical qubits (including ancilla qubits) and the time cost in terms of code cycles, where a code cycle corresponds to measuring all surface-code check operators exactly once. These numbers can be multiplied to obtain the space-time cost in qubitcycles. This is a meaningful figure of merit that should be minimized. It is more meaningful than only the space cost or only the time cost, since distillation protocols can be straightforwardly parallelized, using twice as many qubits to distill states twice as fast.

\begin{figure*}[t!]
\centering
\def\svgwidth{0.98\linewidth}
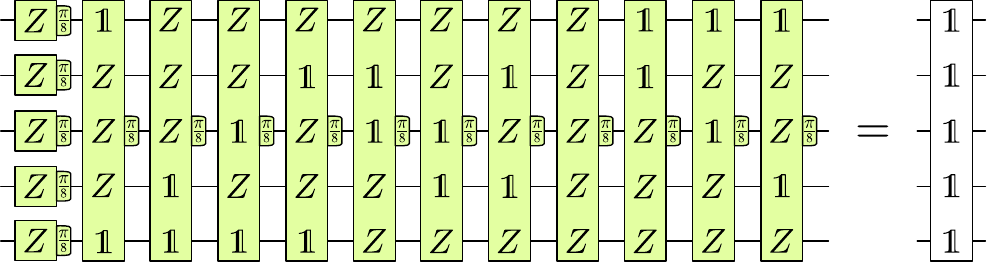
\caption{A sequence of 16 $\pi/8$ rotations on 5 qubits that is non-trivially equivalent to the identity.}
\label{fig:16rotations}
\end{figure*}

Even though this is not necessarily a meaningful quantity, we report the space-time cost in terms of the full distance $d$ for two different choices of $d$ in the last two columns of Tab.~\ref{tab:results}. While the smallest classically intractable quantum computations require ${\sim}100$ qubits, more complicated quantum algorithms use thousands of qubits, such as factoring 2048-bit numbers using Shor's algorithm. The lower and higher values of $d$ are chosen such that they are sufficient for a 100-qubit and 10,000-qubit computation with at most $1/p_{\rm out}$ $T$ gates, respectively. The reported costs in terms of the full distance are in terms of (physical data qubits)$\times$(code cycles), i.e., they do not consider physical measurement ancillas and are therefore smaller by a factor of 2.  This is done to more easily compare the numbers to the cost of storing a $d\times d$ surface-code patch for $d$ code cycles, which is $1d^3$ in terms of (physical data qubits)$\times$(code cycles).

\textbf{How to interpret the cost.} Table~\ref{tab:results} shows protocols that generate one magic state, 4 magic states, or one $\ket{\rm CCZ}$ state that can be used to execute a Toffoli gate. For protocols that generate multiple magic states, the space-time cost and output error are per magic state.
Our protocols feature order-of-magnitude overhead reductions compared to the previous state of the art for all parameter regimes. One example is the $\text{(15-to-1)}_{9,3,3}$ protocol, where the subscripts label the code distances used in the protocol, as explained in Sec.~\ref{sec:15to1}. For $p_{\rm phys}=10^{-4}$, it generates magic states with $p_{\rm out} = 9.3 \times 10^{-10}$, sufficiently low for classically intractable 100-qubit computations with $10^8$ $T$ gates. In a quantum computer that can execute one $T$ gate every $d$ code cycles, 231 logical qubits at $d=13$ would be used to store the 100 qubits with a low error rate~\cite{Litinski2019}, taking into account the routing overhead. A space-time cost of $4.71d^3$ for distillation implies that a footprint equivalent to $4.71$ full-distance qubits would be able to distill one magic state every $d$ code cycles. In this example, ${\approx}2\%$ of the approximately 80,000 physical qubits are used for distillation. The numbers become even more extreme for the example of a 10,000-qubit computation with $10^8$ $T$ gates. Here, the 10,000 data qubits are stored using ${\sim}20{,}000$ logical qubits with $d=15$, which means that the space-time cost of distillation is $3.07d^3$ per magic state. For a quantum computation on more qubits or with a lower overall error probability, distance-17 data qubits might be required, reducing the cost to $2.11d^3$. In this example, the cost to distill a magic state would be lower than the space-time cost of a full-distance logical CNOT gate, which is $3d^3$ per qubit~\cite{Horsman2012}, demonstrating that the cost of magic state distillation is not very high, and that space-time costs that are quantified in units of $d^3$ are of limited usefulness. These numbers are admittedly a bit contrived, but even in the more realistic case of a 100-qubit computation with $p_{\rm phys} = 10^{-3}$ and $p_{\rm out}\approx 10^{-10}$, only ${\approx}10\%$ of all physical qubits are used for distillation. 

\begin{figure*}[t!]
\centering
\def\svgwidth{0.87\linewidth}
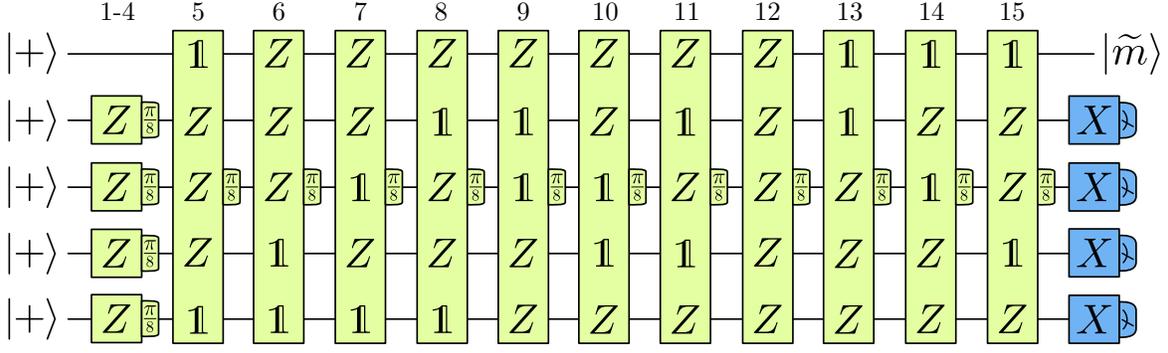
\caption{15-to-1 distillation circuit.}
\label{fig:15to1circuit}
\end{figure*}

\textbf{The main message} is that magic state distillation is \textit{not} the dominant cost in a surface-code-based quantum computer. Rather, the large overhead of surface codes is due to their low encoding rate, which implies that a large number of qubits is required to simply store all data qubits of the computation. 

\textbf{Overview.} In the following sections, we discuss how the protocols in Tab.~\ref{tab:results} are constructed. We start in Sec.~\ref{sec:circuits} by reviewing how distillation circuits work and how their performance is quantified. Distillation protocols require faulty $T$ gates on the level of logical qubits, which are usually performed via state injection and measurement. In Sec.~\ref{sec:faultygates}, we introduce additional protocols for faulty logical $T$ gates based on shrinking patches and faulty $T$ measurements, which avoid Clifford corrections and use fewer qubits and cycles. Next, in Sec.~\ref{sec:15to1}, we go through the construction of the low-cost 15-to-1 protocol. In Sec.~\ref{sec:concatenated}, we construct two-level protocols, where 15-to-1-distillation output states are fed into a second level of 15-to-1 or 20-to-4. In Sec.~\ref{sec:synthillation}, we discuss synthillation protocols, i.e., the distillation of resource states that perform entire layers of $\pi/8$ rotations. Specifically, we show the example of $\ket{CCZ}$ state distillation, which can replace four $T$-gate magic states for the execution of a controlled-controlled-$Z$ gate. For the protocols in Tab.~\ref{tab:results}, the distillation costs of CCZ states are lower than the cost of four $T$-gate magic states with a similar $p_{\rm out}$, indicating that synthillation can lower the cost compared to the distillation of $T$-gate magic states. Finally, in Sec.~\ref{sec:smallfootprint}, we discuss how protocols with a higher space-time cost, but smaller qubit footprint can be constructed. The examples shown in Tab.~\ref{tab:results} reduce the error rate from $10^{-3}$ or $10^{-4}$ to ${\sim}10^{-9}$, but use only as few as 762 or 7,780 physical qubits.

\section{Distillation circuits}
\label{sec:circuits}

Magic state distillation protocols can be understood in terms of quantum error-correcting codes with transversal $T$ gates~\cite{Bravyi2005,Bravyi2012}, but it is conceptually simpler to explain them in terms of circuits~\cite{Campbell2017}. When writing quantum circuits as sequences of Pauli product rotations $P_{\varphi} = e^{-iP\varphi}$, specifically $\pi/8$ rotations $P_{\pi/8}$, certain sequences are equivalent to the identity. While some of these sequences are trivial, e.g., $P_{\pi/8}$ followed by $P_{-\pi/8}$, there also exist non-trivial sequences. One such sequence of 16 rotations on 5 qubits is shown in Fig.~\ref{fig:16rotations}. In general, such sequences are described by triorthogonal matrices~\cite{Bravyi2012,Campbell2017}. The equivalent concept of phase-polynomial identities is used in the context of circuit optimization~\cite{Amy2019}.

If we multiply the circuit in Fig.~\ref{fig:16rotations} by a single-qubit rotation $Z_{-\pi/8}$ on the first qubit, the first rotation will be cancelled and the remaining circuit will consist of 15 rotations, as in Fig.~\ref{fig:15to1circuit}. Since the 16-rotation circuit is equivalent to the identity, the 15-rotation circuit is equivalent to a single $Z_{-\pi/8}$ rotation on the first qubit. In other words, if the initial state is $\ket{+}^{\otimes 5}$, where $\ket{+} = (\ket{0}+\ket{1})/\sqrt{2}$, then the circuit prepares the state \linebreak $\ket{\widetilde{m}}\otimes \ket{+}^{\otimes 4}$. Here, $\ket{\widetilde{m}} = (\ket{0} + e^{-i\pi/4}\ket{1})/\sqrt{2}$ is a state that can be used to perform $\pi/8$ rotations in the same way as $\ket{m}$, but the outcome of the $P \otimes Z$ measurement in Fig.~\ref{fig:magicstateuse} needs to be interpreted differently, i.e., this state is a magic state.

\begin{figure*}[t!]
\centering
\def\svgwidth{\linewidth}
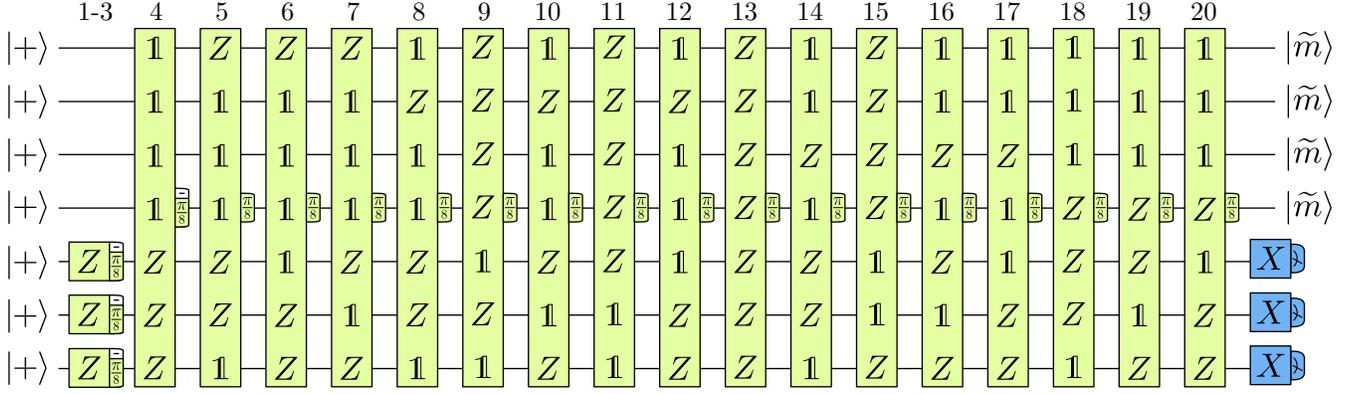
\caption{20-to-4 distillation circuit.}
\label{fig:20to4circuit}
\end{figure*}

Because all rotations in Fig.~\ref{fig:15to1circuit} act non-trivially on qubits 2-5, these qubits can be used to detect errors. If the circuit is executed without errors, qubits 2-5 are initialized in the $\ket{+}$ state and returned to the $\ket{+}$ state, i.e., have an outcome of +1 upon $X$ measurement. Errors are detected, if any of these measurement outcomes are $-1$, in which case the protocol fails and the state is discarded.

The 15-to-1 protocol~\cite{Bravyi2005} is sometimes characterized as having an output error probability of $35p^3$. This assumes that every $P_{\pi/8}$ rotation generates a Pauli error $P=P_{\pi/2}$ with a probability of $p$. Since these are $Z$-type Pauli errors, they will flip all $X$ measurement outcomes of the qubits that they act on. Therefore, any one faulty $P_{\pi/8}$ gate can be detected. Furthermore, there is no combination of two faulty gates that can go undetected. However, some combinations of three faulty gates, e.g., rotations 5, 11 and 14, will cause a $Z$ Pauli error on the output state, but will not trigger any flipped $X$ measurement outcomes. Since there are 35 such combinations, the probability to generate an undetected error is $35p^3$ to leading order.

To compute the subleading corrections to the output error, this process can be simulated numerically. Starting with the initial state $\rho_{\rm init} = \ket{+}\bra{+}^{\otimes 5}$, each of the 15 rotations is applied by mapping
\begin{equation}
	\rho~\rightarrow~(1-p)\cdot P_{\pi/8}^{\vphantom{\dagger}}\rho P_{\pi/8}^\dagger + p \cdot P_{5\pi/8}^{\vphantom{\dagger}}\rho P_{5\pi/8}^\dagger \, .
\end{equation} 
The output state is determined by projecting into the subspace with the correct measurement outcomes using the projectors $\Pi_X = (\mathbbm{1} + X)/2$, i.e.,
\begin{equation}
	\rho_{\rm out} = \frac{1}{1-p_{\rm fail}} (\mathbbm{1} \otimes \Pi_X^{\otimes 4}) \rho (\mathbbm{1} \otimes \Pi_X^{\otimes 4}) \, ,
\end{equation}
where
\begin{equation}
	p_{\rm fail} = 1-\mathrm{tr} \left[  (\mathbbm{1} \otimes \Pi_X^{\otimes 4}) \rho \right] 
\end{equation}
is the failure probability of the protocol. The output error probability is computed by comparing the ideal output state $\rho_{\rm ideal} = \ket{\widetilde{m}}  \bra{\widetilde{m}} \otimes \ket{+}\bra{+}^{\otimes 4}$ to the actual output state $\rho_{\rm out}$. This is done by computing the infidelity
\begin{equation}\begin{split}
	p_{\rm out} &= 1 - F(\rho_{\rm ideal},\rho_{\rm out}) \\
	&= 1 - \mathrm{tr} \left( \sqrt{\sqrt{\rho_{\rm out}} \rho_{\rm ideal} \sqrt{\rho_{\rm out}}} \right)^2 \\
	&= 1- \mathrm{tr} \left( \rho_{\rm ideal} \rho_{\rm out} \right) \, ,
\end{split}\end{equation}
where the last equality holds, because $\rho_{\rm ideal}$ is a pure state. The infidelity corresponds to the probability that a faulty magic state that is used to perform a gate in a quantum circuit will lead to an error of this circuit's outcome~\cite{Haah2017}. Notably, in the examples that we consider, the trace distance $\mathrm{tr}\sqrt{(\rho_{\rm ideal} - \rho_{\rm out})^2}/2$ yields identical or at least similar results. For the example of $p = 10^{-4}$, the approximate output error probability is $35p^3 = 3.5 \times 10^{-11}$, whereas the exact result is $p_{\rm out} = 3.501 \times 10^{-11}$.

\textbf{Random Pauli errors.} If faulty $P_{\pi/8}$ rotations are performed by preparing faulty magic states and using the circuit in Fig.~\ref{fig:magicstateuse}, then the output error depends on the error model for the preparation of faulty magic states. In particular, if the faulty magic state is affected by a random Pauli error with probability $p$, i.e., by an $X$, $Y$ or $Z$ error with probabilities $p/3$, respectively, then this translates into a probability of $p/3$ of performing either a $P_{-\pi/8}$, $P_{3\pi/8}$ or $P_{5\pi/8}$ rotation instead of a $P_{\pi/8}$ rotation. In other words, after each rotation, the state is mapped to
\begin{equation}\begin{split}
	\rho~\rightarrow ~ (1-p) \cdot P_{\pi/8}^{\vphantom{\dagger}}\rho P_{\pi/8}^\dagger &+ \frac{p}{3} \cdot P_{-\pi/8}^{\vphantom{\dagger}}\rho P_{-\pi/8}^\dagger \\
	+ \frac{p}{3} \cdot P_{3\pi/8}^{\vphantom{\dagger}}\rho P_{3\pi/8}^\dagger &+ \frac{p}{3} \cdot P_{5\pi/8}^{\vphantom{\dagger}}\rho P_{5\pi/8}^\dagger \, ,
\label{eqn:errormap}
\end{split}\end{equation} 
so there is a $p/3$ probability of either a $P_{-\pi/4}$, $P_{\pi/4}$ or $P_{\pi/2}$ error. These first two errors are more forgiving than a proper $P_{\pi/2}$ Pauli error, since they effectively only lead to a Pauli error with $50\%$ probability. As a consequence, we expect each of the 35 combinations of three faulty rotations to contribute to the output error with $(8/27)p^3$ instead of $1p^3$: Out of the 27 combinations in $\{ P_{-\pi/4}, P_{\pi/4}, P_{\pi/2}\}^3$, there is one combination with three $P_{\pi/2}$'s, which leads to an undetected error. There are 6 combinations with two $P_{\pi/2}$'s leading to an error with a $50\%$ probability, 20 combinations with one $P_{\pi/2}$ leading to an error with a $25\%$ probability, and 8 combinations with no $P_{\pi/2}$'s, leading to an error with a $12.5\%$ probability. Therefore, the output error should be $p_{\rm out} = 35\cdot \frac{8}{27} p^3 \approx 10.3704p^3$ to leading order. Indeed, a numerical treatment of the full density matrix for $p=10^{-4}$ yields $p_{\rm out} = 1.03724\times 10^{-11}$. 

\textbf{Coherent errors.} The previous two error models randomly applied Pauli errors with a certain probability. One might object that, for physical qubits, this is not necessarily a realistic error model. A more realistic error model would take coherent errors into account, such as systematic under- and over-rotation. Distillation circuits can also detect these errors, but their performance is indeed worse than for incoherent errors. For example, consider the map
\begin{equation}
	\rho~\rightarrow~P_{\pi/8+\varphi}^{\vphantom{\dagger}}\rho P_{\pi/8+\varphi}^\dagger  \, ,
\end{equation} 
which systematically over-rotates each gate by an excess angle $\varphi$. A gate that over-rotates by an angle $\varphi = \arcsin(1/100)$ has the same gate fidelity as a gate that applies a $Z$ error with a probability of $10^{-4}$. However, the infidelity of the output magic state $p_{\rm out} = 1.22 \times 10^{-9}$ is higher by almost two orders of magnitude compared to the incoherent case. In our resource analysis, we will be working with incoherent circuit-level Pauli noise, applying errors according to Eq.~\eqref{eqn:errormap}, but with three different probabilities for the three different errors. Still, we comment on how coherent errors might affect the output error in Sec.~\ref{sec:concatenated}.

\textbf{20-to-4 distillation.} Distillation protocols can output more than one magic state. If the 16-rotation circuit in Fig.~\ref{fig:16rotations} is multiplied by two $Z_{-\pi/8}$ rotations, one on the first and one on the second qubit, a 14-rotation circuit is obtained that outputs a $\ket{\widetilde{m}} \otimes \ket{\widetilde{m}} \otimes \ket{+}^{\otimes 3}$ state, i.e., two magic states. Similarly, using a 24-rotation circuit that non-trivially corresponds to the identity, a 20-rotation circuit that outputs a $\ket{\widetilde{m}}^{\otimes 4} \otimes \ket{+}^{\otimes 3}$ state can be obtained. This is the 20-to-4 protocol~\cite{Bravyi2012} shown in Fig.~\ref{fig:20to4circuit}. With a $Z$-Pauli error model, there are 22 pairs of rotations that can lead to an output error. Therefore, the probability of an output error is $22p^2$ to leading order. However, since four states are produced, one should interpret this as $p_{\rm out} = 5.5p^2$ per magic state. In other words, the probability that the resource state $\ket{\widetilde{m}}^{\otimes 4}$ will cause an error in a circuit is $22p^2$, but, since this resource state executes four $\pi/8$ rotations, this translates into a $5.5p^2$ error probability per gate. In a numerical simulation, the output error per state is determined via the infidelity between the projected output state and the ideal output state $\ket{\widetilde{m}}  \bra{\widetilde{m}}^{\otimes 4} \otimes \ket{+}\bra{+}^{\otimes 3}$  divided by four. For $p=10^{-4}$, this yields $p_{\rm out} = 5.505 \times 10^{-8}$ per output state.

\section{Faulty logical $T$ gates}
\label{sec:faultygates}

\begin{figure}[t!]
\centering
\def\svgwidth{\linewidth}
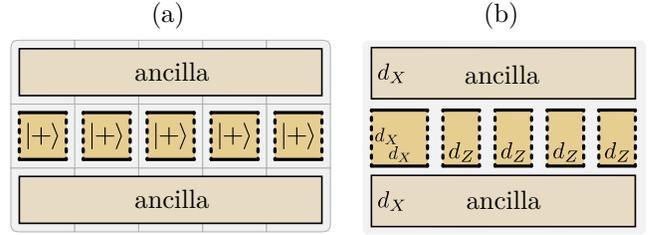
\caption{A naive arrangement (a) of logical qubits could consist of five $d \times d$ patches initialized in the $\ket{+}$ state and two additional $5d \times d$ ancilla regions for Pauli product measurements. The arrangement that we consider (b) consists of one $d_X \times d_X$ patch, four $d_Z \times d_X$ patches, and two ancilla regions with a width $d_X$ for Pauli product measurements.}
\label{fig:15to1arrangement}
\end{figure}

We use the notation of Ref.~\cite{Litinski2019} to draw arrangements of logical surface-code qubits, where patches with dashed and solid edges represent $d \times d$ surface-code patches with $X$ and $Z$ boundaries. Logical operations are performed by measuring products of logical Pauli operators via lattice surgery~\cite{Horsman2012,Litinski2017b,Fowler2018}. A naive layout for the 15-to-1 protocol is shown in Fig.~\ref{fig:15to1arrangement}a, where the five qubits of the 15-to-1 circuit are placed next to each other with their $Z$ boundaries facing up and down. A $5d \times d$ ancillary space above and below these five qubits can be used to measure Pauli product operators between these qubits to perform $\pi/8$ rotations.

The code distance determines the logical error rate of the encoded qubits, which also depends on the underlying error model. Here, we consider circuit-level noise, where each physical gate, state initialization and measurement outcome is affected by a Pauli error with probability $p_{\rm phys}$. Using a minimum-weight perfect matching decoder for such a noise model, the logical error rate per code cycle~\cite{Fowler2018} can be approximated as
\begin{equation}
	p_L(p_{\rm phys},d) = 0.1(100p_{\rm phys})^{(d+1)/2} \, .
\end{equation}
Since a failure to decode $X$ or $Z$ syndromes correctly leads to logical $Z$ or $X$ errors, respectively, we will assume that logical $X$ and $Z$ errors each occur with a probability of $0.5p_L(p_{\rm phys},d)$ per code cycle. 

Not all errors are equally harmful in the context of distillation protocols. Consider $X$ and $Z$ errors that affect one of the five qubits during the 15-to-1 protocol. $Z$ errors affecting the first qubit (i.e., the output qubit) are always detrimental, since they cannot be detected and contribute to the overall output error of the protocol. The effect of $X$ errors on any of the five qubits is to turn all previous $P_{\pi/8}$ rotations that acted on this qubit into $P_{-\pi/8}$ rotations. For instance, consider an $X$ error on the third qubit after rotation 7 in Fig.~\ref{fig:15to1circuit}. This $X$ error can be commuted to the beginning of the circuit and absorbed into the initial $\ket{+}$ state. The commutation turns rotations 2, 5 and 6 into $-\pi/8$ rotations, since $X$ and $Z$ anti-commute. As errors on multiple rotations can lead to undetected errors, $X$ errors should also be avoided.

$Z$ errors on qubits 2-5, on the other hand, are less damaging. They are detectable, as they have the same effect as $Z$ errors that affect rotations 1-4. Therefore, it is not necessary to encode the logical $Z$ operators of qubits 2-5 with the same distance as their logical $X$ operators. Instead, we encode these qubits using rectangular $d_X \times d_Z$ patches with $d_Z \leq d_X$. Their probability of $X$ errors is $0.5(d_Z/d_X)\cdot p_L(p_{\rm phys},d_X)$ per code cycle, since the $X$ distance is $d_X$, but the number of possible $X$ error strings is lower by a factor of $(d_Z/d_X)$ compared to a square patch. Correspondingly, the probability of $Z$ errors is $0.5(d_X/d_Z)\cdot p_L(p_{\rm phys},d_Z)$, since the $Z$ distance is $d_Z$, but the number of $Z$ error strings is higher by a factor of $(d_X/d_Z)$ compared to a square patch. We also fix the distance used in the ancillary region to $d_{X}$. Finally, there is a third distance $d_m$ which determines the number of code cycles used in lattice surgery. This affects the error of the Pauli product measurements used for logical gates, which can be detected by the distillation protocol.

In total, we end up with the arrangement shown in Fig.~\ref{fig:15to1arrangement}b that is characterized by three code distances $d_X$, $d_Z$ and $d_m$, where $d_X$ and $d_Z$ are spatial distances, and $d_m$ is the temporal distance. Before we construct a surface-code implementation of the 15-to-1 protocol, we first discuss two different ways of performing faulty logical $\pi/8$ rotations with surface codes: the traditional method based on state injection, and a protocol based on faulty $T$ measurements.

\subsection{State injection}

The standard method to perform faulty $T$ gates with topological codes is via state injection and measurement. State injection is a protocol that prepares an arbitrary logical state $\ket{\psi_L}$ from a corresponding arbitrary physical state $\ket{\psi}$. Several such state-injection protocols exist~\cite{Fowler2012,Horsman2012,Landahl2014,Li2015,Lodyga2015,Lao2019}, but none of them are fault-tolerant, i.e., the error probability of $\ket{\psi_L}$ is always proportional to $p_{\rm phys}$. The simplest protocol~\cite{Horsman2012} starts with a physical state $\ket{\psi}$, i.e., a $1\times 1$ surface-code patch, and then grows it into a $d \times 1$ patch, and finally into a $d \times d$ patch. This is not a very efficient protocol, since growing patches involves measuring stabilizers for $d$ code cycles. The qubit, therefore, spends many cycles in a distance-1 state, which increases the error probability.

More sophisticated state-injection protocols use post-selection~\cite{Li2015,Lao2019} to decrease the error. If the error rate of single-qubit operations is significantly lower than the error rate of two-qubit gates, the error due to state injection can even be lower than $p_{\rm phys}$. In circuit-level noise, a single number $p_{\rm phys}$ characterizes all gates. However, physical systems typically feature significantly better single-qubit operations than two-qubit gates. In state-of-the-art superconducting-qubit~\cite{SuperconductingQubitGates} and ion-trap~\cite{Wright2019} architectures, for instance, the fidelities of single-qubit and two-qubit gates differ by up to almost two orders of magnitude. Since two-qubit gates are typically the lowest-fidelity operations, and syndrome-readout circuits of surface codes mostly consist of two-qubit gates, the characteristic error rate $p_{\rm phys}$ in circuit-level noise will be largely determined by the error rate of two-qubit gates. If the two-qubit error rate is $p_{\rm phys}$, but the single-qubit error rate is $p_{\rm phys}/10$, state injection can produce magic states with an error as low as $\frac{13}{30}p_{\rm phys}$~\cite{Lao2019} in just two code cycles. However, there is a certain failure rate of the protocol due to post-selection, which increases the length of the protocol.

\begin{figure}[t!]
\centering
\def\svgwidth{0.97\linewidth}
%% Accompanies image file 'injectionvsmeasurement.pdf' (pdf, eps, ps)
\begingroup%
  \makeatletter%
  \providecommand\color[2][]{%
    \errmessage{(Inkscape) Color is used for the text in Inkscape, but the package 'color.sty' is not loaded}%
    \renewcommand\color[2][]{}%
  }%
  \providecommand\transparent[1]{%
    \errmessage{(Inkscape) Transparency is used (non-zero) for the text in Inkscape, but the package 'transparent.sty' is not loaded}%
    \renewcommand\transparent[1]{}%
  }%
  \providecommand\rotatebox[2]{#2}%
  \newcommand*\fsize{\dimexpr\f@size pt\relax}%
  \newcommand*\lineheight[1]{\fontsize{\fsize}{#1\fsize}\selectfont}%
  \ifx\svgwidth\undefined%
    \setlength{\unitlength}{163.97747556bp}%
    \ifx\svgscale\undefined%
      \relax%
    \else%
      \setlength{\unitlength}{\unitlength * \real{\svgscale}}%
    \fi%
  \else%
    \setlength{\unitlength}{\svgwidth}%
  \fi%
  \global\let\svgwidth\undefined%
  \global\let\svgscale\undefined%
  \makeatother%
  \begin{picture}(1,0.43287082)%
    \lineheight{1}%
    \setlength\tabcolsep{0pt}%
    \put(0,0){\includegraphics[width=\unitlength,page=1]{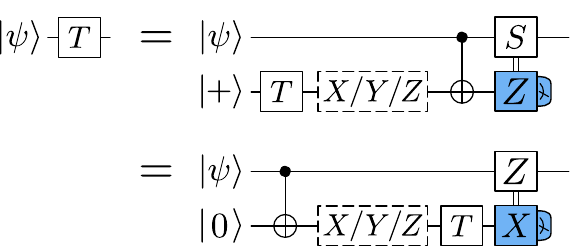}}%
    \put(0.24915262,0.40587475){\color[rgb]{0,0,0}\makebox(0,0)[lt]{\lineheight{1.25}\smash{\begin{tabular}[t]{l}(a)\end{tabular}}}}%
    \put(0.24915262,0.17024294){\color[rgb]{0,0,0}\makebox(0,0)[lt]{\lineheight{1.25}\smash{\begin{tabular}[t]{l}(b)\end{tabular}}}}%
  \end{picture}%
\endgroup%

\caption{A faulty $T$ gate performed via state injection (a) and a faulty $T$ measurement (b).}
\label{fig:injectionvsmeasurement}
\end{figure}

While state injection can be used to prepare faulty magic states, it cannot be used to directly execute $P_{\pi/8}$ rotations. Instead, state injection is used indirectly by preparing a faulty magic state and measuring $P \otimes Z$ via lattice surgery, as shown in Fig.~\ref{fig:magicstateuse}. With a $50\%$ probability, a $P_{\pi/4}$ correction is required. Performing this correction operation either requires extra time or extra space. In any case, this Clifford correction has an effect on the distillation protocol and, therefore, needs to be performed, increasing the space-time cost of the protocol. For this reason, we will avoid state injection, and instead construct a protocol that executes faulty $P_{\pi/8 }$ rotations without the need for Clifford corrections.

\begin{figure}[t!]
\centering
\def\svgwidth{\linewidth}
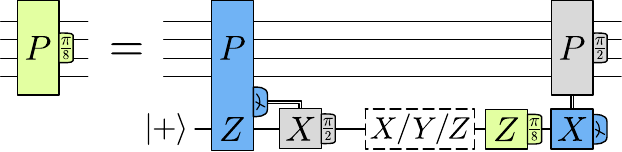
\caption{A faulty $P_{\pi/8}$ rotation corresponds to a $P \otimes Z$ measurement involving a $\ket{+}$ ancilla, followed by a faulty $T$ measurement of the ancilla.}
\label{fig:tmeas}
\end{figure}

\subsection{Faulty $T$ measurements}

\label{sec:slowmeas}

When a $T$ gate is performed on a qubit $\ket{\psi}$ via state injection, a faulty magic state is prepared, entangled with $\ket{\psi}$, and measured, as shown in Fig.~\ref{fig:injectionvsmeasurement}a. The faulty preparation can be treated as a $T$ gate applied on a $\ket{+}$ state followed by a random $X$, $Y$ or $Z$ Pauli error. The idea of faulty $T$ measurements is to avoid the Clifford correction by reversing the order of the entangling operation and the faulty $T$ gate, as shown in Fig.~\ref{fig:injectionvsmeasurement}b. Here, $\ket{\psi}$ is first entangled with a $\ket{0}$ qubit. Next, a sequence of a random Pauli error, a $T$ gate and an $X$ measurement is performed, which we refer to as a \textit{faulty $T$ measurement}. Now, the correction operation in response to the $X$ measurement is no longer a Clifford gate, but a Pauli $Z$ operation, which requires no additional hardware operations. $X$, $Y$ and $Z$ errors lead to $S^\dagger$, $S$ and $Z$ errors on $\ket{\psi}$, respectively. 
Thus, a $P_{\pi/8}$ rotation can be performed by measuring $P \otimes Z$ involving an ancilla qubit initialized in the $\ket{+}$ state, followed by a faulty $T$ measurement of the ancilla qubit, as shown in Fig.~\ref{fig:tmeas}.

\begin{figure}[t!]
\centering
\def\svgwidth{\linewidth}
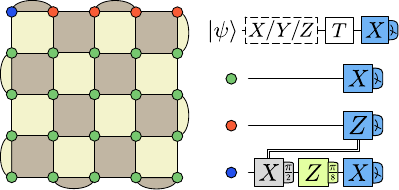
\caption{Surface-code implementation of a faulty $T$ measurement. Bright and dark faces correspond to $Z$-type and $X$-type stabilizers, respectively.}
\label{fig:scfaultymeas}
\end{figure}

With surface codes, protocols for faulty $T$ measurements are exactly identical to protocols for state injection, except that the order of operations is reversed. Here, we describe a simplified protocol to demonstrate the working principle of faulty $T$ measurements. Similarly to the case of state injection, one can construct significantly more sophisticated protocols, as we discuss in Appendix~\ref{appendix}.

%This protocol for faulty $T$ measurements is inspired by a similar protocol for state injection that is currently being developed by Lao \textit{et al.}~in Ref.~\cite{Lao2019}. 

One particularly simple state-injection protocol is performed by growing a physical qubit (a $1 \times 1$ patch) into a $d \times 1$ patch and then into a $d \times d$ patch~\cite{Horsman2012}. The corresponding faulty-$T$-measurement protocol can be performed by shrinking patches. Suppose that a logical qubit $\ket{\psi_L} = \alpha \ket{0_L} + \beta \ket{1_L}$ is encoded in a $d \times d$ patch as in Fig.~\ref{fig:scfaultymeas}, with logical $Z$ operators as strings from left to right. This $d \times d$ patch can be shrunk to a $d \times 1$ patch by measuring all green qubits in the $X$ basis. The remaining $d \times 1$ patch encodes the qubit in a $d$-qubit $XX$ repetition code 
\begin{equation}
\ket{\psi_L} = \frac{\alpha}{\sqrt{2}}(\ket{+}^{\otimes d} + \ket{-}^{\otimes d}) +\frac{\beta}{\sqrt{2}}(\ket{+}^{\otimes d} - \ket{-}^{\otimes d}) \, ,
\end{equation}
where the logical $Z$ operator is $Z^{\otimes d}$, and the logical $X$ operator corresponds to the $X$ operator on any of the $d$ qubits. Next, the $d \times 1$ patch is shrunk to a $1 \times 1$ patch by measuring all red qubits in the $Z$ basis. In fact, the red and green measurements can be performed simultaneously. The product of all $Z$ measurement outcomes (and also preceding stabilizer measurements) determines an $X$ Pauli correction on the remaining qubit, which now stores the logical information in its physical Pauli operators. Finally, a physical $T$ gate is applied to the remaining qubit, before it is measured in the $X$ basis. 
 
 \begin{figure}[t!]
\centering
\def\svgwidth{0.95\linewidth}
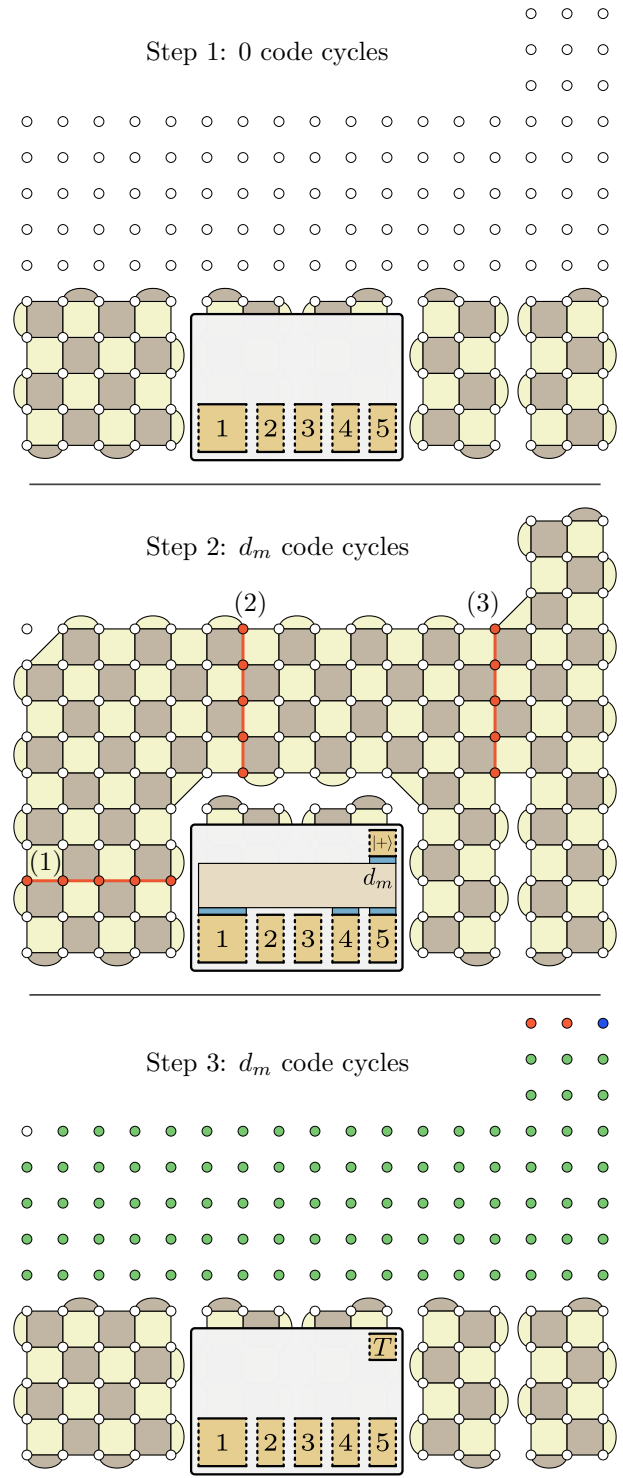
\caption{Example of a faulty $T$ measurement to perform a $(Z_1 \otimes Z_4 \otimes Z_5)_{\pi/8}$ rotation.}
\label{fig:msmt1}
\end{figure}

\begin{figure*}[t!]
\centering
\def\svgwidth{0.95\linewidth}
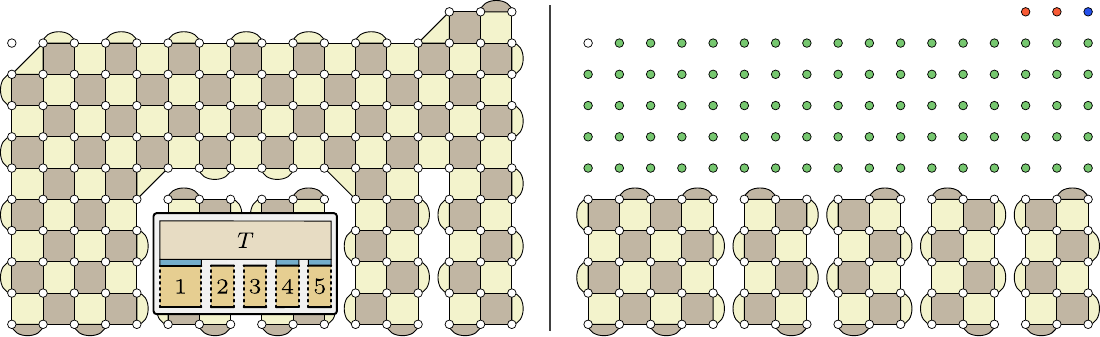
\caption{Shrinking the $\ket{+}$ patch in the top right corner to a $d_m \times 1$ patch produces a stabilizer configuration that is topologically equivalent to the one shown in Fig.~\ref{fig:msmt1}. This does not reduce the code distance compared to Fig.~\ref{fig:msmt1},}
\label{fig:msmt2}
\end{figure*}

Much like state injection, faulty $T$ measurements are not fault-tolerant protocols, in the sense that their error rate is proportional to the physical error rate and does not decrease with the code distance. For a Pauli error model, these error rates can be understood as the probabilities of the Pauli-error operations in the dashed boxes in Fig.~\ref{fig:injectionvsmeasurement}. For simplicity, we will assume that faulty $T$ measurements have a Pauli error rate of $p_{\rm phys}$, meaning that, effectively, the blue qubit is affected by an $X$, $Y$ or $Z$ error with a probability of $p_{\rm phys}/3$ for each Pauli. When used to execute a $P_{\pi/8}$ rotation, this implies that this gate will have a $P_{-\pi/4}$, $P_{\pi/4}$ or $P_{\pi/2}$ error with a probability of $p_{\rm phys}/3$ for each error. This assumption is actually very inaccurate for the protocol shown in Fig.~\ref{fig:scfaultymeas}, since, for this protocol, the error scales with the code distance $d$, as any single-qubit $X$ or $Y$ error on the red qubits translates into a logical error in the faulty $T$ measurement. This can be avoided by changing the measurement pattern, similar to the state-injection protocols of Refs.~\cite{Landahl2014,Lodyga2015}. Moreover, the error rate can be significantly suppressed by adding ``pre-selection'' to the faulty-$T$-measurement protocol, the analogous operation to post-selection in the state-injection protocol of Ref.~\cite{Li2015}, wherein a faulty $T$ measurement is deferred until the stabilizer checks surrounding the sensitive blue qubit report no syndromes for two code cycles. Because such a faulty-$T$-measurement protocol is identical to the state-injection protocol of Ref.~\cite{Li2015}, apart from the reversed order of operations, its error rate can be lower than $p_{\rm phys}$, if single-qubit operations have a significantly higher fidelity than two-qubit operations.

In any case, the specific choice of faulty-$T$-measurement protocol (or state-injection protocol) will not matter for the distillation protocols that we construct in the following sections, which is why the more sophisticated $T$-measurement protocols are discussed in Appendix \ref{appendix}. Moreover, we show in Appendix \ref{appendix} that, for many relevant distillation protocols, a higher error rate for faulty $T$ measurements has only a small effect on the performance of the distillation protocols, even if the error rate of faulty $T$ measurements is assumed to be higher by an order of magnitude, $10p_{\rm phys}$ instead of $p_{\rm phys}$. In our following resource estimates, we will use the simplified assumption of an error rate of $p_{\rm phys}$ for faulty $T$ measurements.

\begin{figure*}[t!]
\centering
\def\svgwidth{0.9\linewidth}
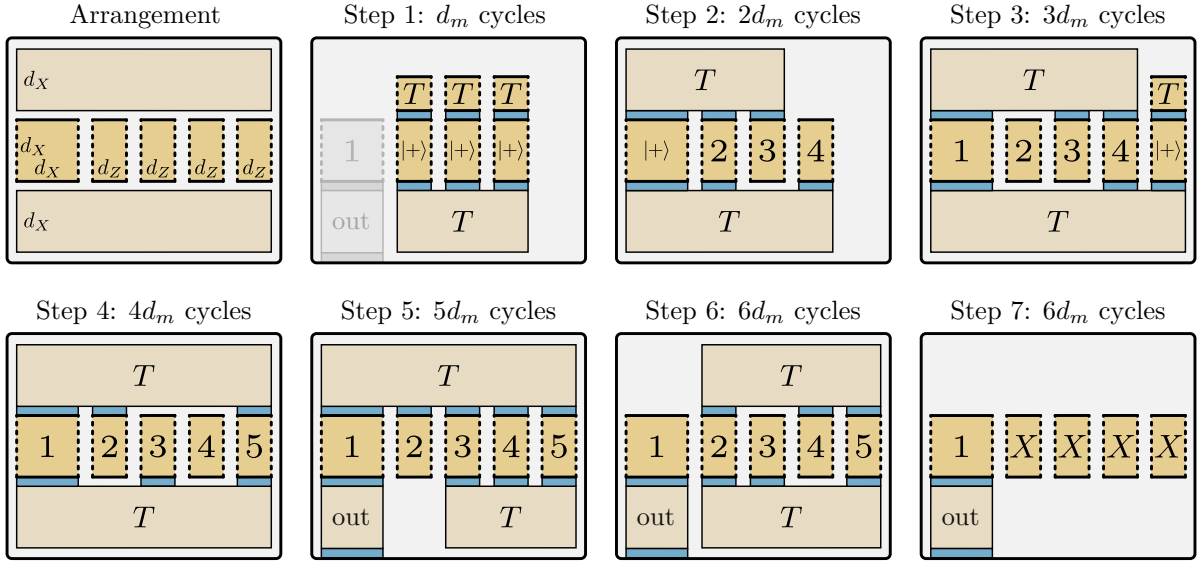
\caption{Surface-code implementation of the $\text{(15-to-1)}_{d_X,d_Z,d_m}$ protocol using $2\cdot (d_X+4d_Z)\cdot 3d_X + 4d_m$ physical qubits for $6d_m$ code cycles.}
\label{fig:15to1impl}
\end{figure*}

We now show how to use faulty $T$ measurements to perform $P_{\pi/8}$ rotations in a surface-code arrangement similar to Fig.~\ref{fig:15to1arrangement}b. Suppose that we want to execute a $P_{\pi/8}=(Z_1 \otimes Z_4 \otimes Z_5)_{\pi/8}$ rotation on three qubits, i.e., rotation 9 in Fig.~\ref{fig:15to1circuit}. In the first step in Fig.~\ref{fig:msmt1}, we start with the five qubits of the 15-to-1 protocol, $(d_X + 4d_Z)\times d_X$ unused qubits in the ancillary region and an additional $d_m \times d_m$ unused qubits in the top right corner. In this example, $d_X =5$, $d_Z=3$ and $d_m=3$. Next, following the circuit in Fig.~\ref{fig:tmeas}, we initialize a $d_m \times d_m$ patch in the $\ket{+}$ state and use multi-patch lattice surgery~\cite{Fowler2018,Litinski2019} to measure $Z_1 \otimes Z_4 \otimes Z_5 \otimes Z_a$, where $Z_i$ is the $Z$ operator of qubit $i$ and $Z_a$ is the $Z$ operator of the ancilla in the top right corner. The multi-patch lattice-surgery operation~\cite{Fowler2018} is performed by initializing the physical qubits in the ancilla region in the $\ket{+}$ state and measuring the new stabilizers for $d_m$ code cycles, after which the qubits in the ancilla region are measured in the $X$ basis. The outcome of the  $Z_1 \otimes Z_4 \otimes Z_5 \otimes Z_a$ measurement is encoded in the measurement outcomes of the $Z$ stabilizers in the ancilla region. Finally, in the last step, a faulty $T$ measurement is performed on the $d_m \times d_m$ patch.

Next, we explain how we perform the error estimate for this protocol. For a protocol that performs a $P_{\pi/8}$ rotation, we consider $Z$ and $X$ storage errors on qubits 1-5 and $P_{\pi/2}$, $P_{\pi/4}$ and $P_{-\pi/4}$ errors in the application of the rotation. As explained previously, the incorrect decoding of $X$ and $Z$ syndromes on qubits 1-5 leads to $Z$ and $X$ storage errors. These occur with a probability of $0.5(d_X/d_H)\cdot p_L(p_{\rm phys},d_H)$ and $0.5(d_H/d_X)\cdot p_L(p_{\rm phys},d_X)$, respectively, where $d_H=d_X$ for qubit 1, and $d_H=d_Z$ for qubits 2-5. The incorrect decoding of the $X$ syndrome in the ancilla region causes $Z$ error strings connecting the $X$ boundaries in the ancilla region. Depending on the exact location of these error strings, this can cause $Z$ errors on one or multiple qubits. For instance, the error string highlighted in red and labeled `(2)' in Fig.~\ref{fig:msmt1} is topologically equivalent to error string `(1)', and therefore causes a $Z$ error on qubit 1. Error string `(3)', on the other hand, is less harmful, since it is equivalent to a $Z_1 \otimes Z_4$ error affecting qubits 1 and 4. Still, as a simplified pessimistic estimate, we will assume that any such error directly contributes to the $Z$ error of the output qubit, i.e., qubit 1. This happens with a probability of $0.5(l/d_X)\cdot p_L(p_{\rm phys},d_X) \cdot d_m$, where $l$ is the length of the ancilla patch. In our example, $l=d_X +4d_Z$.

The incorrect decoding of the $Z$ syndrome in the ancilla region
causes $X$ error strings. While there are no $Z$ boundaries in the ancilla region for the $X$ errors to connect to, in a space-time picture, these errors can also condense at the temporal boundaries set by the initializations of the physical qubits in the $\ket{+}$ state and their measurement in the $X$ basis. Because the lattice-surgery stabilizers are measured for $d_m$ code cycles, the probability of this error is governed by $d_m$. Since $0.5p_L(p_{\rm phys},d_m)\cdot d_m$ is the $X$ error rate of a $d_m \times d_m$ patch stored for $d_m$ code cycles, but the lattice surgery takes place over an area of $l \times d_X$, we estimate the probability of such an error as $0.5 l\cdot d_X/d_m^2 \cdot p_L(p_{\rm phys},d_m)\cdot d_m$. Such an error leads to an incorrect interpretation of the lattice-surgery measurement outcome.  As shown in Fig.~\ref{fig:tmeas}, incorrectly interpreting the outcome of the $P \otimes Z$ measurement causes an $X$ error on the $\ket{+}$ qubit, which turns the $Z_{\pi/8}$ rotation into a $Z_{-\pi/8}$ rotation, causing an on overall $P_{-\pi/4}$ error.

$Z$ and $X$ storage errors affecting the $\ket{+}$ qubit in the top right corner each occur with a probability of $0.5 p_L(p_{\rm phys},d_m) \cdot d_m$ and cause $P_{\pi/2}$ and $P_{-\pi/4}$ errors, respectively, as explained by inserting $X$ and $Z$ errors on the $\ket{+}$ qubit in Fig.~\ref{fig:tmeas}. Finally, the single-qubit $X$, $Y$ and $Z$ errors with a probability of $p_{\rm phys}/3$ during the faulty $T$ measurement of Fig.~\ref{fig:scfaultymeas} contribute to the $P_{-\pi/4}$, $P_{\pi/4}$ and $P_{\pi/2}$ error, respectively.

In order to save qubits, we can shrink the $d_m \times d_m$ patch in the top right corner to a $ d_m \times 1$ patch. As shown in Fig.~\ref{fig:msmt2}, this produces a stabilizer configuration that is topologically equivalent to the configuration of Fig.~\ref{fig:msmt1} and maintains the code distance. In principle, the $Z$ storage error on the $\ket{+}$ qubit is now reduced, but we will still use the error estimate explained in the previous paragraph.

To summarize our error estimate, in addition to the usual storage errors, a $Z$ error affects the output qubit with a probability of $0.5(l/d_X)\cdot p_L(p_{\rm phys},d_X) \cdot d_m$. \linebreak $P_{\pi/2}$, $P_{-\pi/4}$ and $P_{\pi/4}$ errors each occur with a probability of $p_{\rm phys}/3$. Furthermore, a $P_{-\pi/4}$ error occurs with a probability of $0.5 l\cdot d_X/d_m^2 \cdot p_L(p_{\rm phys},d_m)\cdot d_m$. Moreover, additional $P_{-\pi/4}$ and $P_{\pi/2}$ errors occur with a probability of $0.5 p_L(p_{\rm phys},d_m) \cdot d_m$. Since we are not running an actual decoder to obtain these error probabilities, our error estimate should not be interpreted as a rigorous simulation, but as a pessimistic ballpark estimate.

\section{15-to-1 distillation}
\label{sec:15to1}

Our implementation of the 15-to-1 protocol of Fig.~\ref{fig:15to1circuit} is shown in Fig.~\ref{fig:15to1impl}. It starts by initializing qubits 2-4 in the $\ket{+}$ state. In the first step, rotations 1-3 and 5 are performed. The single-qubit rotations are performed by initializing a $d_Z \times d_m$ surface-code patch in the $\ket{+}$ state, measuring $Z \otimes Z$, and performing faulty $T$ measurement of the $d_Z \times d_m$ patch. Multi-qubit rotations are performed via fast faulty $T$ measurements. In step 2, qubit 1 is initialized in the $\ket{+}$ state and rotations 6 and 7 are performed. In step 3, qubit 5 is initialized in the $\ket{+}$ state and rotations 4, 8 and 9 are performed. In steps 4-6, rotations 10-15 are performed. Finally, in step 7, qubits 2-5 are measured in the $X$ basis. If all measurement outcomes are +1, qubit 1 is a distilled magic state which can be used to execute a low-error $P_{\pi/8}$ rotation. The distillation block can now be used to distill the next magic state.

In order to prevent the output state from blocking the space reserved for qubit 1 for $d_X$ code cycles, the consumption of the output state for the execution of a $P_{\pi/8}$ rotation can already be initiated in step 5, since this process takes $d_X$ code cycles. In the protocol shown in Fig.~\ref{fig:15to1impl}, steps 5-7 and step 1 of the subsequent distillation can be used to measure qubit 1, i.e., a total of $3d_m$ code cycles. If $d_X > 3d_m$, the output state will block the space reserved for qubit 1 and slow down the protocol. This can be prevented by reordering the rotations. In any case, we only consider cases with $d_X \leq 3d_m$.

\begin{figure}[t!]
\centering
\def\svgwidth{0.95\linewidth}
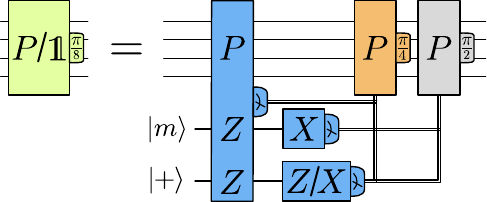
\caption{Quantum circuit for delayed-choice $P_{\pi/8}$ rotations. The choice of measurement basis for the single-qubit $\ket{+}$ measurement decides whether a $P_{\pi/8}$ rotation is performed, or no operation at all.}
\label{fig:delayedchoice}
\end{figure}

One may be concerned that, when consuming the output state in step 5, we still do not know if the distillation protocol will succeed, i.e., whether the output state is faulty or not. This problem can be solved by using the circuit in Fig.~\ref{fig:delayedchoice}, where an additional ancilla qubit initialized in the $\ket{+}$ state is used. To execute a $P_{\pi/8}$ rotation, the operator $P \otimes Z \otimes Z$ between the qubits, the magic state and the additional ancilla state is measured. Depending on whether the $\ket{+}$ state is measured in the $Z$ or $X$ basis, the $P_{\pi/8}$ rotation is applied or not. This can be used to consume a magic state before the outcome of its distillation is known. If the distillation protocol for this magic state fails, the $\ket{+}$ state can be measured in the $X$ basis to prevent the faulty magic state from being used.

In total, this  15-to-1 protocol has a space cost of $2 \cdot (d_X + 4d_Z)\cdot 3d_X + 4d_m$ physical qubits, taking physical measurement ancillas into account. The time cost is $6d_{m}/(1-p_{\rm fail})$ code cycles, where $p_{\rm fail}$ is the failure probability of the protocol. As discussed in Sec.~\ref{sec:circuits}, $p_{\rm fail}$ and the output error $p_{\rm out}$ are determined numerically. To this end, the 5-qubit density matrix is simulated, taking into account errors from storage and faulty $T$ measurements. All $P_{\pi/8}$ rotations have a $P_{\pi/2}$, $P_{\pi/4}$ and $P_{-\pi/4}$ error of $p_{\rm phys}/3$ and additional errors due to the fast faulty $T$ measurement protocol, as discussed in Sec.~\ref{sec:faultygates}, with the exception that, for rotations that do not involve qubit 1, $Z$ errors in the ancilla region do not cause $Z$ errors on qubit 1. For single-qubit rotations, $X$ errors on the $d_Z \times d_m$ qubit cause $P_{-\pi/4}$ errors and occur with a probability of $0.5d_Z p_L(p_{\rm phys},d_m)$, whereas $Z$ errors spread to the adjacent qubit with a probability of $0.5(d_m^2/d_Z)p_L(p_{\rm phys},d_Z)$. 

After each step (apart from step 7), $d_m$ code cycles worth of $X$ and $Z$ storage errors are applied to all five qubits. In addition, $Z$ or $X$ errors on the $d_X \times d_X$ ancilla patch used to consume the output state are added as $Z$ or $X$ storage errors to the output state for $d_X$ code cycles. Finally, the output error probability $p_{\rm out}$ is computed as the infidelity $1- F(\rho_{\rm ideal}, \rho_{\rm out})$.

\begin{figure*}[t!]
\centering
\def\svgwidth{\linewidth}
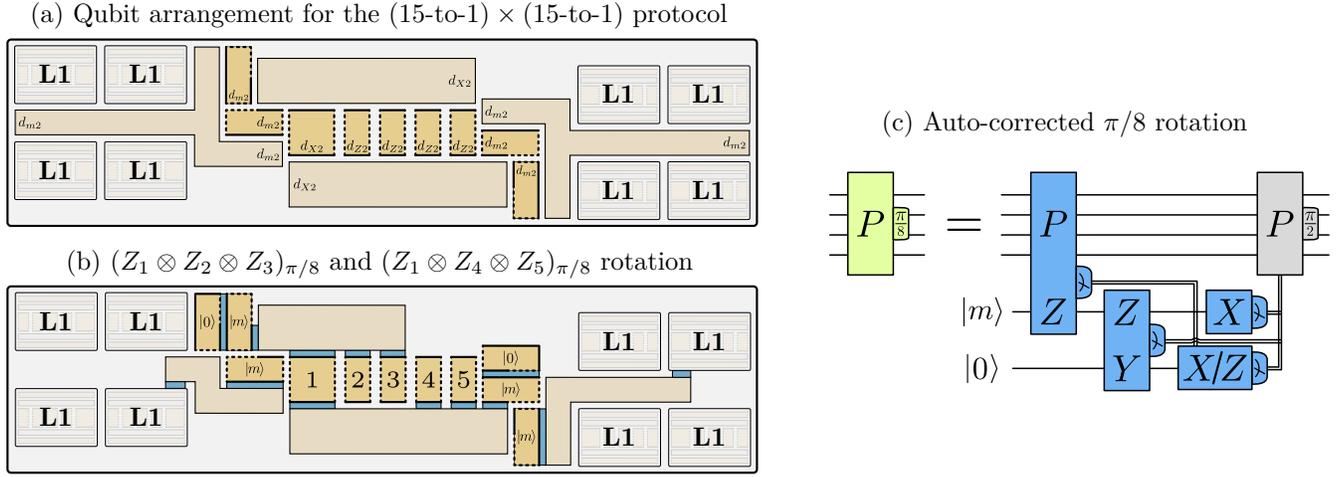
\caption{(a) Example of a qubit arrangement used for two-level $\text{(15-to-1)}^{n_{L1}}_{d_X,d_Z,d_m} \times \text{(15-to-1)}_{d_{X2},d_{Z2},d_{m2}}$ distillation with $n_{L1}=8$ level-1 blocks. (b) Example of two rotations performed in this arrangement. (c) The rotations are performed using auto-corrected $\pi/8$ rotations, which consume a magic state and apply the Clifford correction by appropriately choosing the measurement basis of the $\ket{0}$ ancilla qubit.}
\label{fig:twolevel15to1layout}
\end{figure*}

We refer to a 15-to-1 protocol characterized by the code distances $d_X$, $d_Z$ and $d_m$ as a $\text{(15-to-1)}_{d_X,d_Z,d_m}$ protocol. For $p_{\rm phys}=10^{-4}$, we find that the protocol $\text{(15-to-1)}_{7,3,3}$ has a $p_{\rm out} = 4.4 \times 10^{-8}$, where we round the output error to two significant digits. It has a space cost of 810 qubits and a time cost of 18.1 code cycles. These two numbers can be multiplied to obtain a space-time cost of 14,600 qubitcycles to three significant digits. We also report the space-time cost in terms of the \textit{full distance} $d_{\rm full}$. Consider a 100-qubit quantum computation with a $T$-gate count of $n_T$, where $n_T < 1/p_{\rm out}$. Using the construction from Ref.~\cite{Litinski2019}, the entire computation can be finished in $n_T \cdot d_{\rm full}$ code cycles, if 231 distance-$d_{\rm full}$ surface-code patches are used to store the 100 qubits. The probability that any of these qubits is affected by a storage error that can spoil the outcome of the computation is 
\begin{equation}
	p_{\text{storage error}} = 231\cdot n_T \cdot d_{\rm full} \cdot p_L(p_{\rm phys}, d_{\rm full}) \, .
\end{equation}
The storage error can be higher, if the computation does not exclusively consist of $\pi/8$ rotations, but also involves Pauli product measurements, as this increases the length of the computation. Using magic states with $p_{\rm out}$ for each $T$ gate, there is a probability of $n_T \cdot p_{\rm out}$ that a faulty $T$ gate will spoil the outcome of the computation. If we demand that storage errors leads to a relative increase of the error probability by $1\%$ in the best case, then $d=d_{\rm full}$ needs to satisfy
\begin{equation}
	231\cdot n_T \cdot d \cdot p_L(p_{\rm phys}, d) < 0.01 \cdot n_T \cdot p_{\rm out} \, .
\end{equation}
Evidently, $d$ does not depend on $n_T$, but only on the number of qubits in the computation. For a 10,000-qubit computation, the qubits need to be stored using 20,284 distance-$d$ qubits, and the condition changes to 
\begin{equation}
	20284\cdot d \cdot p_L(p_{\rm phys}, d) < 0.01 \cdot p_{\rm out} \, .
\end{equation}
We pick the smallest odd-integer $d$ that satisfies the condition. For $p_{\rm out} = 4.4 \times 10^{-8}$, $d = 11$ in the 100-qubit case and $d = 13$ in the 10,000-qubit case. The space-time cost reported in the last two columns of Tab.~\ref{tab:results} is in terms of (physical data qubits)$\times$(code cycles), i.e., $5.49d^3$ or $3.33d^3$. These numbers are remarkably low, as the cost to explicitly perform, e.g., a logical CNOT gate on two distance-$d$ qubits~\cite{Horsman2012} is $3d^3$. Still, the only truly meaningful number quantifying the space-time cost is the cost in terms of qubitcycles.

The other protocols reported in Tab.~\ref{tab:results} for  $p_{\rm phys}=10^{-4}$ are $\text{(15-to-1)}_{9,3,3}$ and $\text{(15-to-1)}_{11,5,5}$, reducing the error to $p_{\rm out} = 9.3 \times 10^{-10}$ and $p_{\rm out} = 1.9 \times 10^{-11}$, respectively. For a higher physical error rate of $p_{\rm phys}=10^{-3}$, $\text{(15-to-1)}_{17,7,7}$ has a $p_{\rm out} = 4.5 \times 10^{-8}$. Interested readers can verify these numbers using the Python script or Mathematica notebook provided in the Supplementary Material~\cite{magicstatesuppl}, where they can also try out other parameters. The Supplementary Material contains the resource cost calculations for all protocols considered in this paper.

15-to-1 protocols cannot generate arbitrarily good output states, as $p_{\rm out}$ is limited by ${\sim}10p_{\rm phys}^3$. In order to distill higher-fidelity magic states, we turn to two-level protocols.

\section{Two-level protocols}
\label{sec:concatenated}

The idea of two-level protocols is to use distilled magic states (level-1 states) to perform a second round of distillation. We first discuss $\text{(15-to-1)} \times \text{(15-to-1)}$ protocols, where 15-to-1 output states are used for a second level of 15-to-1 distillation. The qubit arrangement used for this protocol is shown in Fig.~\ref{fig:twolevel15to1layout}a. It is described by three additional code distances $d_{X2}$, $d_{Z2}$ and $d_{m2}$, and the number of level-1 distillation blocks $n_{L1}$, where $n_{L1}$ is an even integer. The central region consists of a $(d_{X2} + 4d_{Z2}) \times 3d_{X2} $ array of qubits where the second level of distillation takes place. To the left and right are the level-1 distillation blocks that feed level-1 states into the upper and lower ancilla region of the level-2 block, respectively. Each of these two level-1 region consists of $n_{L1}/2$ level-1 blocks. These blocks are characterized by the level-1 distances $d_X$, $d_Z$ and $d_m$, such that each level-1 region produces one magic state every $6d_m/(1-p_{\rm fail})/(n_{L1}/2)$ code cycles. 

In addition, each level-1 region has a $3d_{m2} \times 4d_{m2}$ array of qubits that separates the level-1 blocks from the level-2 block. As level-1 states are produced, they are transported into this intermediate region. For this purpose, each level-1 region has two spots that are reserved for level-1 output states. While one of these spots is being filled with a newly generated level-1 state, the magic state in the other spot can be consumed to execute a $P_{\pi/8}$ rotation in the level-2 block, as shown in Fig.~\ref{fig:twolevel15to1layout}b. These rotations are performed using the auto-corrected $\pi/8$ rotation~\cite{Litinski2019} shown in Fig.~\ref{fig:twolevel15to1layout}c. Here, the operator $P \otimes Z$ between the level-2 qubits and a level-1 magic state $\ket{m}$ is measured. Simultaneously, $Z \otimes Y$ between $\ket{m}$ and an ancillary $\ket{0}$ state is measured. Depending on the outcome of the $P \otimes Z$ measurement, the $\ket{0}$ qubit is either read out in the $X$ or $Z$ basis, essentially performing a Clifford correction or not.

\begin{figure*}[t!]
\centering
\def\svgwidth{0.99\linewidth}
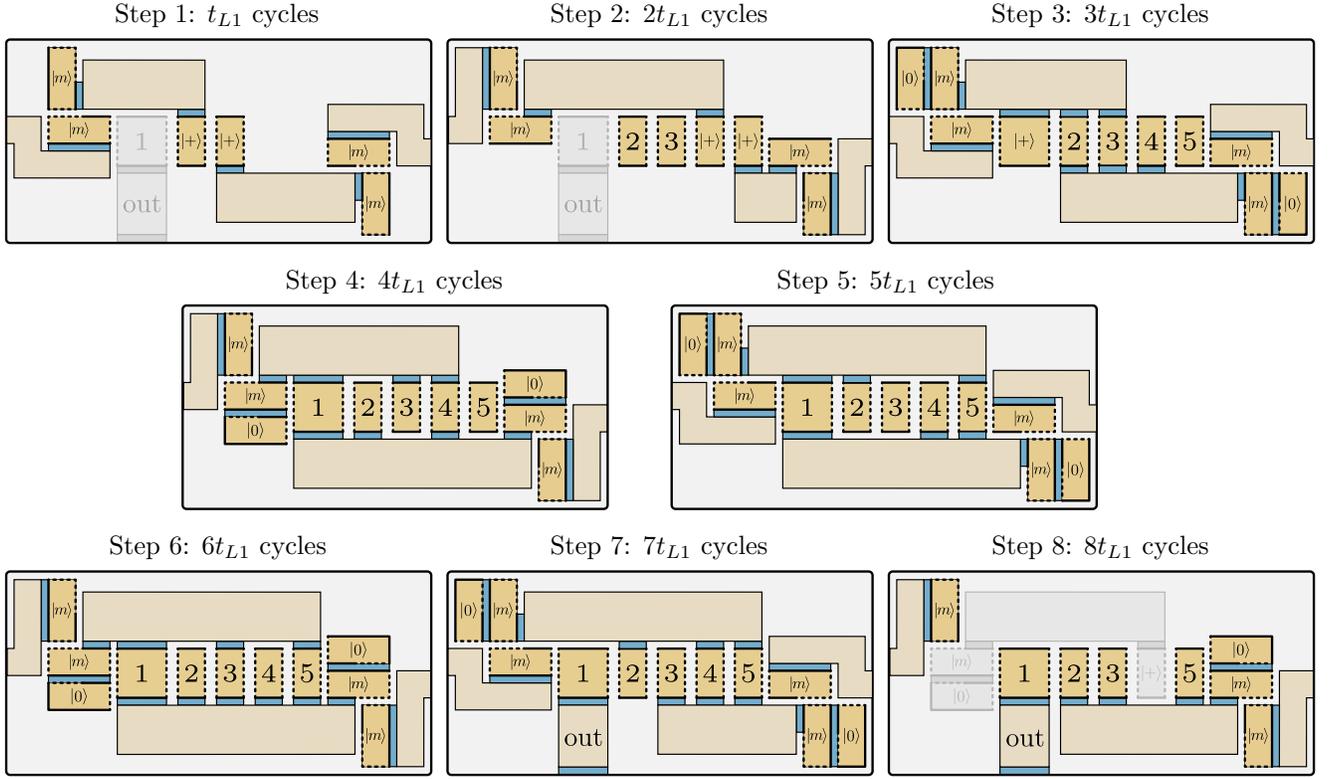
\caption{Surface-code implementation of the  $\text{(15-to-1)} \times \text{(15-to-1)}$ protocol.}
\label{fig:twolevel15to1}
\end{figure*}

As shown in Fig.~\ref{fig:twolevel15to1layout}b, this kind of measurement can be performed in $d_{m2}$ code cycles using a configuration similar to a faulty $T$ measurement. While one magic state in each level-1 region is used to execute a level-2 $\pi/8$ rotation, a second level-1 state is being transported to be used for the subsequent level-2 rotation. In the top and bottom ancilla region, one such level-2 rotation can be performed every
\begin{equation}
  t_{L1} = \max(d_{m2},6d_m/(1-p_{\rm fail})/(n_{L1}/2))
\end{equation}
code cycles. If level-1 states are produced slowly, $t_{L1}$ is determined by the output rate of the level-1 factories. If these states are produced fast, $t_{L1}$ will be limited by $d_{m2}$, the duration of the auto-corrected $\pi/8$ rotations.

The entire $\text{(15-to-1)} \times \text{(15-to-1)}$ protocol is shown in Fig~\ref{fig:twolevel15to1}, focusing on the level-2 region. In steps 1 and 2, the first four rotations of the 15-to-1 circuit are performed. Since these are single-qubit rotations, $\ket{0}$ ancillas are not required, as Clifford corrections correspond to Pauli corrections for these first four rotations. In steps 3-7, rotations 5-14 are performed. The consumption of the output state is initiated in step 7. In step 8, qubit 4 is measured in the $X$ basis and rotation 15 is performed in the bottom ancilla region. Since the space reserved for qubit 4 is now empty, the top ancilla region can be used to perform the first rotation of the subsequent distillation round, such that the next round of distillation will only take $7t_{L1}$ instead of $8t_{L1}$ code cycles. For this reason, the time cost of this distillation block is $7.5t_{L1}$ code cycles. Again, the consumption of the output state will slow down the protocol, if it takes longer than $3d_{m2}$ code cycles, so the distances should be chosen such that $d_{X2} \leq 3d_{m2}$.

\begin{figure*}[t!]
\centering
\def\svgwidth{0.98\linewidth}
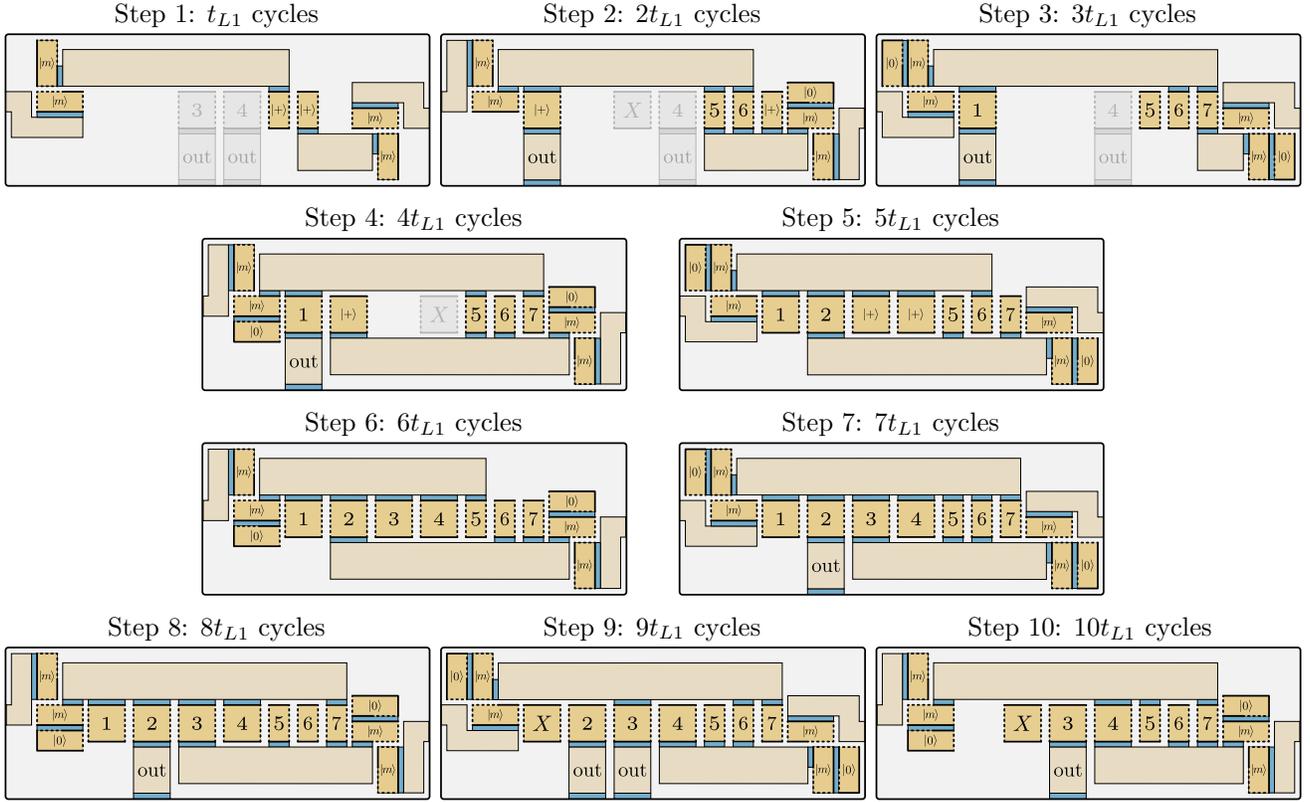
\caption{Surface-code implementation of the  $\text{(15-to-1)} \times \text{(20-to-4)}$ protocol.}
\label{fig:twolevel20to4}
\end{figure*}

\textbf{Error analysis.} The level-1 blocks output level-1 states with an output error of $p_{\rm out} = p_{L1}$. This error contributes to the probability of a $P_{\pi/2}$ error, when this state is used for a $P_{\pi/8}$ rotation. Furthermore, the level-1 state accumulates additional errors as it is moved to the level-2 block. As shown in Fig.~\ref{fig:twolevel15to1layout}b, it traverses a region of width $d_{m2}$ and a maximum length of $n_{L1}/4\cdot (d_X + 4d_Z) + 2d_{m2}$ for $d_{m2}$ code cycles, before ending up in a $2d_{m2} \times 2d_{m2}$ region, where it stays for another $d_{m2}$ code cycles. Therefore, we define
\begin{equation}
  l_{\rm move} = n_{L1}/4\cdot (d_X + 4d_Z) + 10d_{m2} \, ,
\end{equation}
as the length of the ancilla region that increases the storage error of the level-1 state. In this sense, the $X$ and $Z$ error of the level-1 state are each increased by $0.5 l_{\rm move} \cdot p_L(p_{\rm phys},d_{m2})$, contributing to the $P_{\pi/2}$ and $P_{-\pi/4}$ error, respectively. The error analysis of the level-2 block is analogous to faulty $T$ measurements. For an ancilla of length $l$, where $l$ can be up to $d_{X2} + 4d_{Z2} + d_{m2}$, $X$ errors lead to a $P_{-\pi/4}$ error with a probability of $0.5(l\cdot d_{X2}/d_{m2}) \cdot p_L(p_{\rm phys},d_{m2})$. Moreover, if qubit 1 is part of the rotation, it is affected by an additional $Z$ storage error with a probability of $0.5(l\cdot d_{m2}/d_{X2}) \cdot p_L(p_{\rm phys}, d_{X2})$. Finally, the $X$ and $Z$ error of the output state increases by $d_{X2}\cdot p_L(p_{\rm phys},d_{X2})$ while it is being consumed.

\begin{figure*}[t!]
\centering
\def\svgwidth{0.98\linewidth}
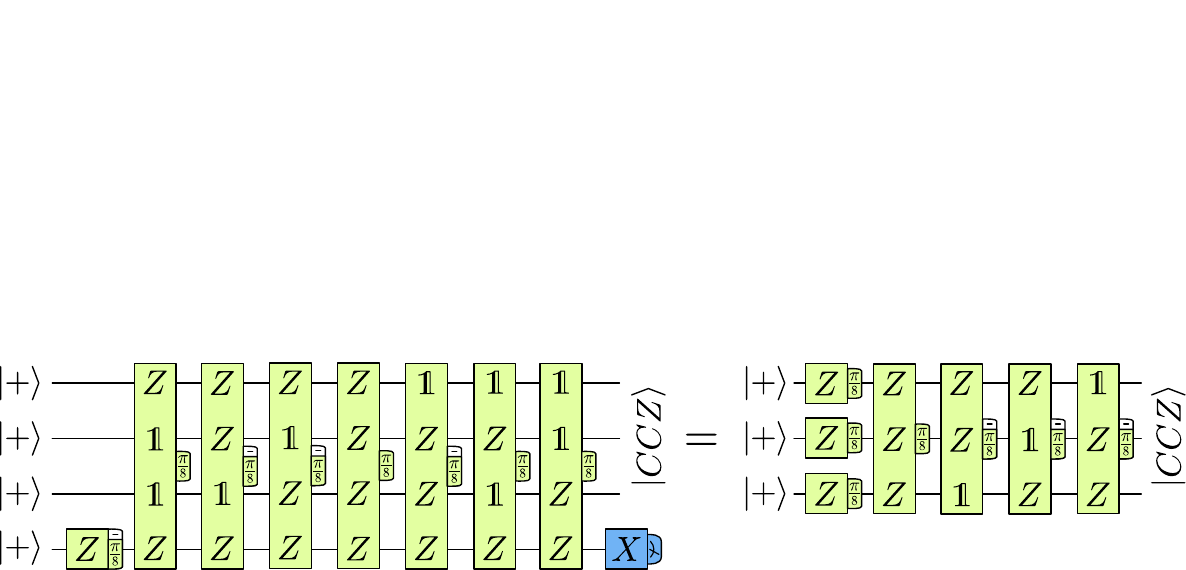
\caption{(a) These 15 rotations are non-trivially equivalent to the identity. (b) 8-to-CCZ synthillation circuit obtained by multiplying the circuit in (a) with a 7-rotation CCZ gate.}
\label{fig:8toCCZcircuit}
\end{figure*}

\textbf{Results.} The space cost of the protocols has three contributions: $2 \cdot (d_{X2} + 4d_{Z2}) \cdot 3d_{X2}$ qubits from the level-2 block, $2n_{L1}((d_{X} + 4d_{Z})(3d_{X}+d_{m2}/2) + 2d_m)$ qubits from the level-1 blocks, including the ancilla that they are connected to, and $2\cdot(20d_{m2}^2 + 2d_{X2}d_{m2})$ qubits from the additional ancilla qubits. Note that, in contrast to the 15-to-1 factory, the footprint of the (15-to-1)$\times$(15-to-1) factory is not necessarily rectangular. The time cost is $7.5t_{L1}$. Applying errors numerically according to our error analysis, we obtain the output error. We label our protocols as $\text{(15-to-1)}^{n_{L1}}_{d_X,d_Z,d_m} \times \text{(15-to-1)}_{d_{X2},d_{Z2},d_{m2}}$. For $p_{\rm phys} = 10^{-4}$, we find that a $\text{(15-to-1)}^4_{9,3,3} \times \text{(15-to-1)}_{25,9,9}$ protocol produces output states with $p_{\rm out} = 6.3 \times 10^{-25}$ for 1,260,000 qubitcycles. For $p_{\rm phys} = 10^{-3}$, the protocol  $\text{(15-to-1)}^6_{11,5,5} \times \text{(15-to-1)}_{29,11,13}$ produces magic states with $p_{\rm out} = 3.3 \times 10^{-14}$ for 3,810,000 qubitcycles. More protocols can be found in Tab.~\ref{tab:results} or generated using the Python script or Mathematica notebook provided in the Supplementary Material~\cite{magicstatesuppl}.

\textbf{20-to-4 distillation.} For output states with a desired error rate that is lower than what is possible with one level of 15-to-1, but higher than what can be achieved with two levels of 15-to-1, it can be more efficient to use a level-2 protocol that is cheaper, but features less error suppression. One such protocol is the 20-to-4 protocol of Fig.~\ref{fig:20to4circuit}. The implementation of a (15-to-1)$\times$(20-to-4) is shown in Fig.~\ref{fig:twolevel20to4} and is very similar to a (15-to-1)$\times$(15-to-1) protocol. The main difference is that the length of the level-2 region increases to $4d_{X2}+3d_{Z2}$, as the 20-to-4 circuit acts on seven qubits, four of which are output states.

For protocols that generate multiple output states, it is particularly important to pick a suitable order in which the rotations are performed, in order to avoid congestion. If four output states are generated at the end of the protocol, but the computation demands that they are consumed one after the other, then they will block the level-2 factory for many cycles. In step 1 of our protocol, rotations 1 and 2 are performed. In step 2, qubit 1 is initialized in the $\ket{+}$ state and rotations 4 and 5 are performed. Simultaneously, the measurement of qubit 1 as an output state can be initiated. In steps 3-6, rotations 3 and 6-12 are performed. In step 7, rotations 13 and 14 are performed and the consumption of output state 2 can be initiated. The snapshots in Fig.~\ref{fig:twolevel20to4} are drawn in a way that assumes that it takes $3t_{L1}$ to consume a magic state, but this depends on $d_{X2}$. In steps 8-9, rotations 15-18 are performed and the consumption of qubit 3 can be initiated. In step 10, the last two rotations are performed and the protocol finishes after $10t_{L1}$ code cycles. The first three steps of the subsequent round of distillation can be used to initiate the measurement of output state 4 and finish the measurements of all remaining output states. If output states are consumed one after the other, e.g., to perform $\pi/8$ rotations one after the other, this protocol allocates up to $2.5t_{L1}$ code cycles for each output state, which is sufficient for the parameters that we consider.

The error analysis of protocols labeled as $\text{(15-to-1)}^{n_{L1}}_{d_X,d_Z,d_m} \times \text{(20-to-4)}_{d_{X2},d_{Z2},d_{m2}}$ is identical to two-level 15-to-1 protocols, albeit with a different space and time cost, and the extra step of dividing the space-time cost and output error by four. Moreover, if multiple output qubits are part of a rotation, the additional $Z$ error due to $Z$-error strings in the ancilla region with a probability of $0.5(l\cdot d_{m2}/d_{X2}) \cdot p_L(p_{\rm phys}, d_{X2})$ is applied to all output qubits that are part of the rotation. For $p_{\rm phys} = 10^{-4}$, we find that the $\text{(15-to-1)}^4_{9,3,3} \times \text{(20-to-4)}_{15,7,9}$ protocol generates states with an error of $p_{\rm out} = 2.4 \times 10^{-15}$ per output state and a space-time cost of 371,000 qubitcycles per magic state. For $p_{\rm phys} = 10^{-3}$, the protocols  $\text{(15-to-1)}^6_{13,5,5} \times \text{(20-to-4)}_{21,11,13}$ and  $\text{(15-to-1)}^4_{13,5,5} \times \text{(20-to-4)}_{27,13,15}$ have output errors of $1.4 \times 10^{-10}$ and $2.6 \times 10^{-11}$ per state with a space-time cost of 1,410,000 and 1,840,000 qubitcycles per output state.

\textbf{What about coherent errors?} As discussed in Sec.~\ref{sec:circuits}, the performance of distillation protocols can be worse, if the underlying error model features coherent errors. In a circuit-level study of surface codes, coherent errors are difficult to analyze. If the logical error rate of surface codes can be maintained at $p_L(p_{\rm phys},d)$ even in the presence of coherent errors, and the only effect of coherent errors is to under- or over-rotate the physical $T$ gate used in faulty $T$ measurements, then one can argue that the effect of coherent errors is not too significant. These errors would then increase the minimum achievable output error rate of the first level of distillation, albeit by an error that is governed by the single-qubit error rate. While this can be a problem for single-level distillation schemes, this is less detrimental for two-level distillation schemes, as only the first level is affected. Since level-1 blocks typically output states that have a fidelity that is much lower than the maximum achievable level-1 fidelity, the overall output fidelity would be barely affected. For instance, the level-1 block of the $\text{(15-to-1)}^4_{13,5,5} \times \text{(20-to-4)}_{27,13,15}$ protocol outputs states with $p_{\rm out} \approx 10^{-6}$, whereas the lowest possible error of the 15-to-1 protocol for $p_{\rm phys} = 10^{-3}$ is  $p_{\rm out} \approx 10^{-8}$. Still, a more careful treatment of coherent errors is necessary, but is beyond the scope of this work.

\textbf{What about feed-forward?} One possible limiting factor in quantum computers is the feed-forward time, i.e., the time it takes to react to measurement outcomes, deciding which operation should be performed next. In our protocols, some qubits need to be measured in the X or Z basis depending on previous measurement outcomes, which is used to avoid Clifford corrections. In Fig.~\ref{fig:twolevel15to1layout}a, these are the qubits in the intermediate region between the level-1 blocks and the level-2 block. If the feed-forward time is a bottleneck in a given architecture, these qubits need to be stored for some additional time before being read out. A slowdown due to feed-forward can be avoided by using additional ancilla qubits in this ancilla region. In any case, long feed-forward times increase the overall space-time cost.

The constructions discussed in the previous sections can be used to implement any distillation protocol that can be expressed as a sequence of $Z$-type $\pi/8$ rotations, e.g., all protocols that are based on triorthogonal matrices~\cite{Bravyi2012,Haah2018}. A very similar class of protocols are synthillation protocols, whose implementation we discuss in the following section.

\section{Synthillation}
\label{sec:synthillation}

Synthillation~\cite{Campbell2017,Campbell2017a} is a portmanteau of the words (gate) \textit{synthesis} and \textit{distillation}. The idea of synthillation is to generate resource states that do not execute single $T$ gates or $\pi/8$ rotations, but entire layers of commuting $\pi/8$ rotations. The simplest example is the $\ket{CCZ}$ resource state, which is prepared by applying a controlled-controlled-$Z$ (CCZ) gate to a $\ket{+}^{\otimes 3}$ state. This state can be used to perform a CCZ gate, which can be written as a sequence of seven commuting $\pi/8$ rotations~\cite{Selinger2013,Litinski2019}. However, it is also possible to execute CCZ gates using four $T$ gates~\cite{Jones2013}, or even only two $T$ gates, if these CCZ gates are part of a compute-uncompute circuit~\cite{Gidney2018}.

\begin{figure}[t!]
\centering
\def\svgwidth{0.8\linewidth}
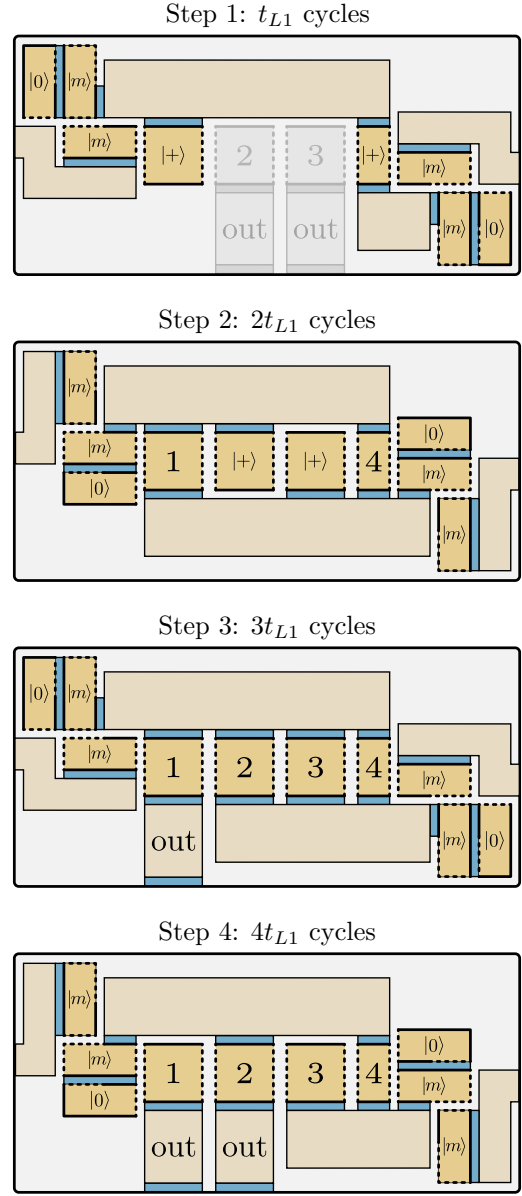
\caption{Surface-code implementation of the $\text{(15-to-1)} \times \text{(8-to-CCZ)}$ protocol.}
\label{fig:8toCCZ}
\end{figure}

Synthillation circuits can be obtained the same way as ordinary distillation circuits. We start with a non-trivial representation of the identity in Fig.~\ref{fig:8toCCZcircuit}a as a sequence of 15 $\pi/8$ rotations on 4 qubits. Next, we cancel the first three and last four rotations by multiplying the entire circuit with the corresponding $+\pi/8$ and $-\pi/8$ rotations, i.e., the 7 rotations on the right-hand side of Fig.~\ref{fig:8toCCZcircuit}b. These 7 rotations happen to correspond to the decomposition of the CCZ gate into 7 $\pi/8$ rotations. In other words, the circuit on the left-hand side of Fig.~\ref{fig:8toCCZcircuit}b prepares a $\ket{CCZ}$ state on the first three qubits and acts trivially on the fourth qubit. Therefore, the fourth qubit can be used to detect errors. If any one of the 8 rotations fails, the $X$ measurement outcome will flip. However, any pair of errors will go undetected. Therefore, the output error to leading order is $28p^2$ under $Z$-Pauli noise.

Since the 8-to-CCZ circuit is a sequence of $Z$-type $\pi/8$ rotations, it can be implemented the same way as a two-level distillation protocol, as shown in Fig.~\ref{fig:8toCCZ}. The level-2 block has a width of $3d_{X2}+d_{Z2}$. Performing two rotations every $t_{L1}$ code cycles, the protocol finishes after $4t_{L1}$ code cycles. In our numerical error analysis, we obtain the output error as the infidelity between the output state and the ideal state $\rho_{\rm ideal} = \ket{CCZ}\bra{CCZ} \otimes \ket{+} \bra{+}$. Labelling our protocols as $\text{(15-to-1)}^{n_{L1}}_{d_X,d_Z,d_m} \times \text{(8-to-CCZ)}_{d_{X2},d_{Z2},d_{m2}}$, we find that, for $p_{\rm phys}=10^{-3}$, the protocol $\text{(15-to-1)}^6_{13,7,7} \times \text{(8-to-CCZ)}_{25,15,15}$ generates $\ket{CCZ}$ states with a fidelity of $p_{\rm out} = 5.2 \times 10^{-11}$ and a space-time cost of 2,820,000 qubitcycles. If the execution of a CCZ gate using ordinary magic states requires four of these states, they need to have a quarter of this output error to achieve the same gate error. In comparison, a $\text{(15-to-1)}\times\text{(20-to-4)}$ protocol can generate one $T$-gate magic state with $p_{\rm out} = 2.6 \times 10^{-11}$ using 1,840,000 qubitcycles per state. In other words, four such states would cost 7,360,000 qubitcycles, more than twice as expensive as a $\ket{CCZ}$ state.

While this implies that synthillation protocols can reduce the distillation cost, this ignores the fact that the consumption of the output state can congest the distillation block. In the construction of Fig.~\ref{fig:8toCCZ}, there are $2t_{L1}$ code cycles for the consumption of each state. If all states can be consumed simultaneously, then this might be sufficient. However, if the rest of the quantum computer is such, that the three output states need to be consumed one after the other, the synthillation protocol may be slowed down significantly, increasing the overall space-time cost. This can be avoided by using additional qubits to temporarily store the output state, although this does not prevent an increase in the space-time cost. Another possibility is to slow down the protocol to increase the time available for the consumption of the output states, e.g, by using slow measurements instead of fast measurements.

For $p_{\rm phys} = 10^{-4}$, we find that a $\text{(15-to-1)}^4_{7,3,3} \times \text{(8-to-CCZ)}_{15,7,9}$ generates output states with $p_{\rm out} = 7.2 \times 10^{-14}$ using 447,000 qubitcycles, which is also cheaper compared to the cost to distill ordinary magic states with the same fidelity. Such synthillation protocols can be used to generate resource states that execute any arbitrary sequence of $\pi/8$ rotations. Schemes to obtain such protocols are found in Ref.~\cite{Campbell2017}. However, note that the problem of output states congesting the distillation block becomes more severe, if the generated output state consists of many qubits.

\section{Small-footprint protocols}
\label{sec:smallfootprint}

So far, our protocols have focused on minimizing the space-time cost. In this section, we outline how protocols can be designed to minimize qubit footprint, i.e., the space cost, by sacrificing space-time overhead. We discuss the example of (15-to-1) and  $\text{(15-to-1)}\times\text{(15-to-1)}$ protocols with a target output error of $p_{\rm out} \approx 10^{-9}$.

\begin{figure}[t!]
\centering
\def\svgwidth{0.6\linewidth}
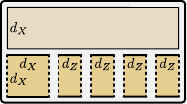
\caption{Qubit arrangement of a small-footprint implementation of the 15-to-1 protocol. }
\label{fig:smallfootprint}
\end{figure}

\textbf{15-to-1.} The footprint can be straightforwardly reduced by using only one region of ancilla qubits instead of two. This reduces the footprint to $4(d_X+4d_Z)d_X+2d_m$ physical qubits, as shown in Fig.~\ref{fig:smallfootprint}. Since only once ancilla region is used, the time cost doubles compared to the protocol of Fig.~\ref{fig:15to1impl} to $12d_m$. The error estimate is identical to the ordinary 15-to-1 protocol. We find that, for $p_{\rm phys}=10^{-4}$, the small-footprint $\text{(15-to-1)}_{9,3,3}$ protocol produces magic states with $p_{\rm out} = 1.5 \times 10^{-9}$ using 762 qubits. However, since the protocol takes 36.2 code cycles, the space-time cost of 27,600 qubitcycles is higher than for comparable protocols with a similar output error.

\begin{figure}[t!]
\centering
\def\svgwidth{0.8\linewidth}
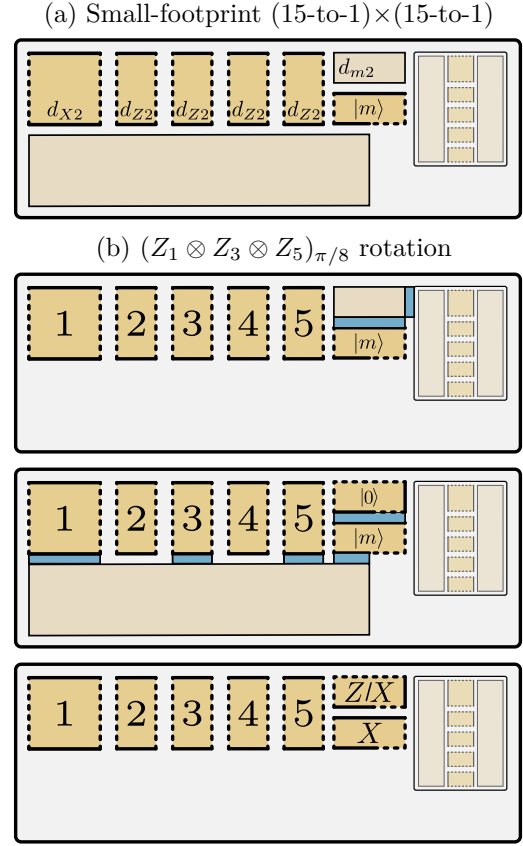
\caption{(a) Qubit arrangement for the small-footprint implementation of the (15-to-1)$\times$(15-to-1) protocol. (b) Example of a $(Z_1 \otimes Z_3 \otimes Z_5)_{\pi/8}$ rotation executed in $2d_{m2}$ cycles.}
\label{fig:smallfootprint2}
\end{figure}

\textbf{Two-level 15-to-1 distillation.} For two levels of 15-to-1 with a small footprint, we use the arrangement shown in Fig.~\ref{fig:smallfootprint2}a, which consists of a $(d_{X2} + 4d_{Z2}) \times 2d_{X2} $ level-2 block, a $(d_X + 4d_Z) \times 3d_X$ level-1 block, and an intermediate region of size $2d_{m2} \times 2d_{m2} + d_{m2} \times d_{X2}$. The level-1 block outputs level-1 states every ${\sim}6d_m$ code cycles. In principle, it is possible to use the small-footprint level-1 block of Fig.~\ref{fig:smallfootprint}, but in the interest of not increasing the space-time cost by too much, we use the 15-to-1 block introduced in Sec.~\ref{sec:15to1}. When a level-1 state is generated, it can be consumed for a level-2 rotation in $2d_{m2}$ code cycles. An example of a $(Z_1 \otimes Z_3 \otimes Z_5)_{\pi/8}$ is shown in Fig.~\ref{fig:smallfootprint2}b. In the first $d_{m2}$ code cycles, a level-1 state is transported from the level-1 block to the intermediate region. In the second $d_{m2}$ code cycles, a Pauli product measurement is performed, executing a $P_{\pi/8}$ rotation according to the circuit in Fig.~\ref{fig:twolevel15to1layout}c. Transporting the level-1 state into the intermediate region increases its $X$ and $Z$ error by $5d_{m2} \cdot p_L(p_{\rm phys},d_{m2})$. If we define the time that it takes to execute a level-2 rotation as
\begin{equation}
  t_{L1} = \max(2d_{m2},6d_m/(1-p_{\rm fail})) \, ,
\end{equation}
then the distillation protocol takes $15t_{L1}$ to finish.

In our numerical analysis, we find that, for $p_{\rm phys} = 10^{-3}$, a small-footprint $\text{(15-to-1)}_{9,5,5} \times \text{(15-to-1)}_{21,9,11}$ protocol generates magic states with $p_{\rm out} = 6.1 \times 10^{-10}$ using only 7,780 physical qubits. With a time cost of 469 cycles, the space-time cost is 3,650,000 qubitcycles per output state. While the space cost is very low, this protocol sacrifices space-time cost, as an ordinary $\text{(15-to-1)}\times\text{(20-to-4)}$ protocol can generate states with $p_{\rm out} = 1.4 \times 10^{-10}$ for only 1,420,000 qubitcycles.

\section{Conclusion}
\label{sec:conclusion}

We have constructed magic state distillation protocols that reduce the space-time cost by approximately $90\%$ compared to the previous state of the art. Since our results were not obtained by simulating entire surface-code patches and running an actual decoder, but via a careful error analysis, the numbers reported in Tab.~\ref{tab:results} should be taken with a grain of salt. The protocols discussed in this paper should rather be regarded as a proof of principle, demonstrating that the overhead of distillation can be reduced significantly by carefully tuning the code distances of the different qubits that are part of the distillation protocol. In any case, exact numbers will have a strong dependence on the hardware-specific error parameters and the decoding procedure.

There is still plenty of room for optimization. For one, we only considered very simple distillation protocols, i.e., the 15-to-1 distillation protocol, a $\text{20-to-4}$ block-code protocol and the synthillation of $\ket{CCZ}$ states. Perhaps, more sophisticated distillation circuits can further reduce the cost. While the 20-to-4 protocol is part of an entire family of $(3k+8)$-to-$k$ protocols, it seems unlikely that a higher $k$ will decrease the cost, since the space cost is governed by $k+3$, and the time cost is governed by $3k+8$. Therefore, the space-time cost per output state is governed by $(3k+8)(k+3)/k$, which happens to have minima at $k=2$ and $k=4$ for even integer $k$. Still, since it is possible to generate arbitrarily many such protocols based on triorthogonal codes~\cite{Bravyi2012,Haah2018}, one could look for protocols that minimize the space-time cost in this manner. For two-level distillation, it remains unclear which combination of protocols reduces the cost. It also remains unclear whether the space-time cost can be decreased by using protocols that reduce the number of $\pi/8$ rotation in the distillation circuit by adding Clifford gates~\cite{Campbell2016}, or protocols that employ catalyst states that need to be stored at a higher code distance~\cite{Gidney2018a}. One could also construct protocols with more than two levels of distillation.

The resource requirements for fault-tolerant surface-code-based quantum computing can be daunting. Hopefully, this work helps demonstrate that this is not mainly due to the overhead of magic-state distillation, but rather due to the low encoding rate of topological codes, implying that thousands of physical qubits are required to encode a single logical qubit.

\section*{Acknowledgments}

I would like to thank Earl Campbell and Craig Gidney for discussions about the advantages of two-level distillation protocols, Lingling Lao for discussions about state injection, and Markus Kesselring for discussions about the surface-code implementations. This work has been supported by the Deutsche Forschungsgemeinschaft (Bonn) within the network CRC TR 183.

\bibliographystyle{apsrev4-1mod}
\bibliography{biblio}

\appendix

\begin{figure*}
\centering
\def\svgwidth{0.99\linewidth}
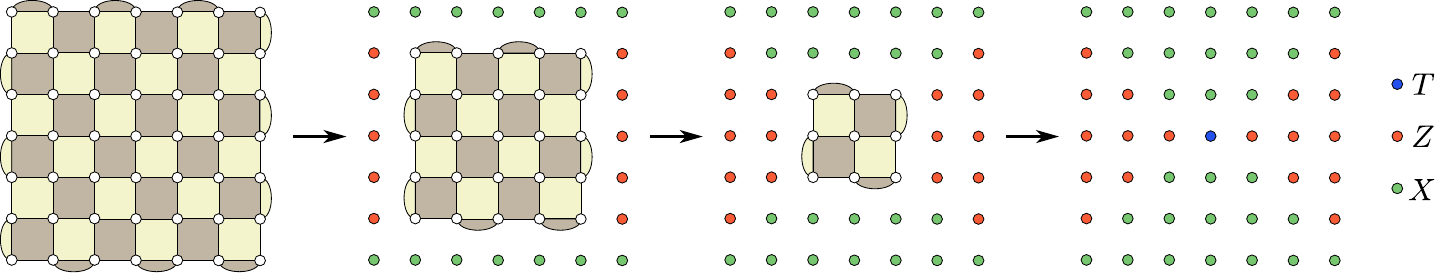
\caption{Protocol of a faulty $T$ measurement in the spirit of Ref.~\cite{Lodyga2015}. It can be thought of as a $d \times d$ patch being shrunk to a $(d-2)\times (d-2)$ patch, then a $(d-4)\times (d-4)$ patch, and so on, until a physical $T$ gate is performed on the remaining $1 \times 1$ patch (the blue qubit), before it is measured in the $X$ basis. Since all these shrinking operations can be performed simultaneously, the $T$-measurement protocol corresponds to just the last step of the figure, where all red qubits are measured in the $Z$ basis, all green qubits in the $X$ basis, and a faulty $T$ measurement is performed on the blue qubit.}
\label{fig:msmt}
\end{figure*}

\section{Faulty $T$ measurements}
\label{appendix}

This appendix discusses faulty $T$ measurements in more detail. Specifically, we discuss approaches to decrease their error rate to justify the use of a faulty-$T$-measurement error rate of $p_{\rm phys}$ in the main text and the potential implications of a higher error rate.

The $T$-measurement protocol discussed in Sec.~\ref{sec:slowmeas} is simple, but the operation of shrinking a $d \times d$ patch to a $d \times 1$ patch is associated with an error that scales with the code distance $d$. While this might not be problematic for small code distances, it implies that this specific $T$-measurement protocol will not work for larger code distances. A more advanced version of this protocol is shown in Fig.~\ref{fig:msmt}. This protocol essentially corresponds to the state-injection protocol of Ref.~\cite{Lodyga2015} in reverse. A $d \times d$ patch is shrunk to a $(d-2) \times (d-2)$ patch by measuring physical qubits along the boundary in the $X$ or $Z$ basis. Next, the patch is shrunk to a $(d-4) \times (d-4)$ patch, a $(d-6) \times (d-6)$ patch, etc., until we end up with a $1 \times 1$ patch, i.e., a single physical qubit corresponding to the logical qubit. This qubit is then measured in the $T$ basis. Since the qubit is in a distance-1 state only at the very end of the protocol, the error rate of this $T$-measurement protocol is not proportional to the code distance $d$.

The shrinking operations can all be performed simultaneously, such that the faulty $T$ measurement corresponds to measuring all red qubits in the last step of Fig.~\ref{fig:msmt} in the $Z$ basis, and all green qubits in the $X$ basis. A physical $T$ gate is applied to the remaining blue qubit, after which it is measured in the $X$ basis. This protocol is almost identical to the $T$-measurement protocol of Sec.~\ref{sec:slowmeas}, with the main difference being the position of the blue qubit. In the protocol of Sec.~\ref{sec:slowmeas}, the blue qubit is in the corner of the patch, whereas in Fig.~\ref{fig:msmt}, the blue qubit is in the center of the patch.

\begin{figure}[b!]
\centering
\def\svgwidth{0.75\linewidth}
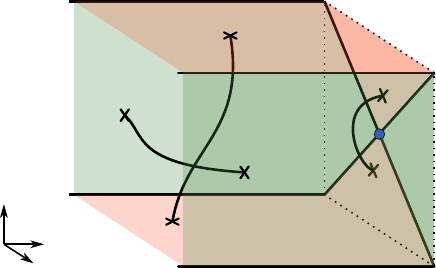
\caption{Space-time diagram of the $T$-measurement protocol of Fig.~\ref{fig:msmt} with time increasing to the right, and four examples of error strings that cause logical errors.}
\label{fig:spacetime}
\end{figure}

In order to examine the error sources of this protocol, we consider the space-time diagram of the $T$-measurement protocol of Fig.~\ref{fig:msmt}, which is shown in Fig.~\ref{fig:spacetime}. It is identical to the space-time diagram of the state-injection protocol of Ref.~\cite{Lodyga2015} shown in Ref.~\cite{Brown2018}. Such space-time diagrams are useful to visualize the effect of physical Pauli errors and stabilizer measurement errors on the encoded logical information. Strings of errors connecting the red boundaries in the space-time diagram lead to logical $Z$ errors, whereas error strings connecting the green boundaries lead to logical $X$ errors. In our usual estimates of idling distance-$d$ logical qubits, we take into account the weight-$d$ error strings that precede the faulty $T$ measurement, i.e., error strings similar to the ones labelled (1) and (2) in Fig.~\ref{fig:spacetime}, which have a probability of $p_L(p_{\rm phys},d)$. During state injection or faulty $T$ measurements, there are some additional low-weight errors corresponding to error strings that connect two time-like boundaries during the $T$ measurement, i.e., single-qubit errors on the blue qubit (3) and strings similar to the one labelled (4) in Fig.~\ref{fig:spacetime}.

\begin{figure*}
\centering
\def\svgwidth{0.8\linewidth}
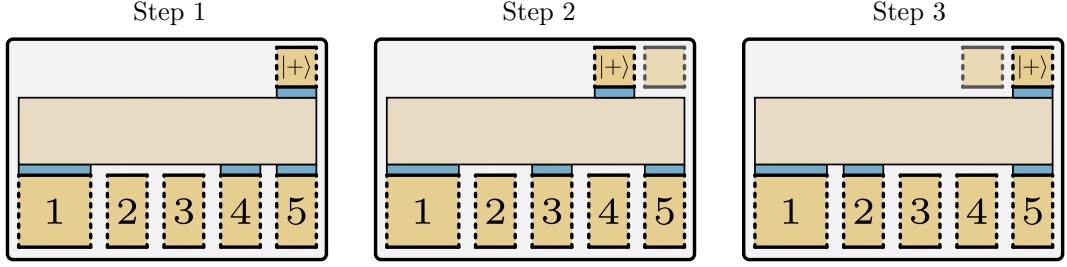
\caption{The finite rejection rate of pre-selection can be prevented from significantly increasing the time cost of distillation protocols by using an additional $d_m \times d_m$ patch during the execution of $P_{\pi/8}$ rotations. While one $d_m \times d_m$ patch is idling until the pre-selection condition is satisfied~--~i.e., until the stabilizers in a region surrounding the sensitive physical qubit have not reported any syndromes for two code cycles~--~a second $d_m \times d_m$ patch can be used to execute the next $P_{\pi/8}$ rotation.}
\label{fig:msmt3}
\end{figure*}

\textbf{Pre-selection.} These low-weight errors are the dominant contribution to the error rate of such state-injection of faulty-$T$-measurement protocols. Specifically, for circuit-level Pauli noise, single-qubit errors affecting the qubits close to the blue qubit and failures of two-qubit gates involved in the syndrome readout of check operators close to the blue qubit contribute to these low-weight errors. In the case of state-injection protocols, these errors can be suppressed through the addition of post-selection~\cite{Li2015}, particularly, if the error rate of two-qubit gates is significantly higher than the error rate of single-qubit operations. An injected magic state is only accepted, if the stabilizer checks in a certain region around the sensitive, blue qubit report no syndrome for two code cycles. Since at least two circuit-level two-qubit errors are required to generate an unreported error, the dominant contribution to the error rate of the state-injection protocol will be governed by single-qubit errors on the sensitive physical qubits. However, this post-selection adds a certain failure probability to the state-injection protocol, as some instances of state injection will be rejected. The protocol in Ref.~\cite{Li2015} reduces the error rate of state injection to ${\sim}2p_{\rm phys}/3$ for two-qubit error rates of $p_2 = p_{\rm phys}$ and one-qubit error rates of $p_1 = p_{\rm phys}/10$ with a rejection rate of around 50\% for $p_{\rm phys} = 10^{-3}$. In Ref.~\cite{Li2015}, a distance-7 region around the sensitive qubit was used for post-selection, leading to a high rejection rate. Preliminary numerical results by Lao et al.~\cite{Lao2019} indicate that by post-selecting a distance-3 region around the sensitive qubit, the error rate of state injection still stays below $p_{\rm phys}$, while the rejection rate drops to ${\sim}4\%$ for $p_{\rm phys} = 10^{-3}$.

\begin{table*}[t]
\centering
\scalebox{0.91}{
\begin{tabular}{c|c|l|c|c|c|c}

\multirow{2}{*}{Protocol} & \multirow{2}{*}{$p_{\rm phys}$} & \multicolumn{1}{c|}{\multirow{2}{*}{$p_{\rm out}$}} & \multirow{2}{*}{Qubits} & \multirow{2}{*}{Cycles} & \multicolumn{2}{c}{Space-time cost per output state} \\
& & & & & Qubitcycles & Full distance \\
\cline{1-7}

$\text{(15-to-1)}_{9,3,3}$ & $10^{-4}$ & $2.1 \times 10^{-8}$ & 1,150 & 18.2 & 20,900 & 
$4.75d^3 \hspace{0.1em} / \hspace{0.1em} d=13$ \vline~$3.10d^3 \hspace{0.1em} / \hspace{0.1em} d=15$   \\

$\text{(15-to-1)}_{7,3,3}^6 \times \text{(20-to-4)}_{13,5,7}$ & $10^{-4}$ & $1.4 \times 10^{-12}$ & 13,200 & 70.0 & 231,000 & $23.5d^3 \hspace{0.1em} / \hspace{0.1em} d=17$ \vline~$16.9d^3 \hspace{0.1em} / \hspace{0.1em} d=19$  \\

$\text{(15-to-1)}_{9,3,3}^4 \times \text{(20-to-4)}_{15,7,9}$ & $10^{-4}$ & $6.6 \times 10^{-15}$ & 16,400 & 91.2 & 374,000 & $27.3d^3 \hspace{0.1em} / \hspace{0.1em} d=19$ \vline~$20.2d^3 \hspace{0.1em} / \hspace{0.1em} d=21$  \\

$\text{(15-to-1)}_{9,3,3}^4 \times \text{(15-to-1)}_{25,9,9}$ & $10^{-4}$ & $4.2 \times 10^{-22}$ & 18,600 & 68.4 & 1,270,000 & 
$32.4d^3 \hspace{0.1em} / \hspace{0.1em} d=27$ \vline~$26.1d^3 \hspace{0.1em} / \hspace{0.1em} d=29$ \\

\cline{1-7}

$\text{(15-to-1)}_{13,5,5}^6 \times \text{(20-to-4)}_{21,11,13}$ & $10^{-3}$ & $5.7 \times 10^{-9}$ & 40,700 & 130 & 1,325,000 & $33.7d^3 \hspace{0.1em} / \hspace{0.1em} d=27$ \vline~$22.2d^3 \hspace{0.1em} / \hspace{0.1em} d=31$ \\

$\text{(15-to-1)}_{11,5,5}^6 \times \text{(15-to-1)}_{21,9,11}$ & $10^{-3}$ & $2.1 \times 10^{-10}$ & 27,400 & 85.7 & 2,350,000 & $48.1d^3 \hspace{0.1em} / \hspace{0.1em} d=29$ \vline~$32.7d^3 \hspace{0.1em} / \hspace{0.1em} d=33$ \\

$\text{(15-to-1)}_{11,5,5}^6 \times \text{(15-to-1)}_{23,11,11}$ & $10^{-3}$ & $2.5 \times 10^{-11}$ & 29,500 & 85.7 & 2,530,000 & $42.5d^3 \hspace{0.1em} / \hspace{0.1em} d=31$ \vline~$29.5d^3 \hspace{0.1em} / \hspace{0.1em} d=35$ \\

$\text{(15-to-1)}_{11,5,5}^6 \times \text{(15-to-1)}_{25,11,11}$ & $10^{-3}$ & $6.4 \times 10^{-12}$ & 30,700 & 85.7 & 2,630,000 & $36.7d^3 \hspace{0.1em} / \hspace{0.1em} d=33$ \vline~$26.0d^3 \hspace{0.1em} / \hspace{0.1em} d=37$ \\

$\text{(15-to-1)}_{13,7,7}^8 \times \text{(15-to-1)}_{29,13,13}$ & $10^{-3}$ & $1.5 \times 10^{-13}$ & 52,400 & 97.5 & 5,110,000 & $59.6d^3 \hspace{0.1em} / \hspace{0.1em} d=35$ \vline~$43.1d^3 \hspace{0.1em} / \hspace{0.1em} d=39$ \\

\end{tabular}}

\caption{Selection of distillation protocols under the assumption of a faulty-$T$-measurement error rate of $10p_{\rm phys}$ instead of $1p_{\rm phys}$ as in Tab.~\ref{tab:results}.}
\label{tab:higherr}
\end{table*}

Since faulty $T$ measurements are identical to state injection, apart from the time direction being reversed, the same approach can be used to reduce the error rate of faulty-$T$-measurement protocols. The analogous operation is \textit{pre-selection}: The single-qubit measurements in the protocol of Fig.~\ref{fig:spacetime} are delayed, until the check operators in a region around the blue qubit have reported no syndrome for two consecutive code cycles. This justifies the use of a faulty-$T$-measurement error rate of $p_{\rm phys}$ in the main text, but also adds a time cost to the fault-$T$-measurement protocol through the rejection rate. While one might be worried that, particularly for high $p_{\rm phys}$, the finite rejection rate in this faulty-$T$-measurement protocol increases the time it takes to execute a faulty $P_{\pi/8}$ rotation, this does not need to be the case, if a few extra qubits are added to the protocol, as shown in Fig.~\ref{fig:msmt3}. Here, in step 1, a $P_{\pi/8}$ rotation is performed using the protocol of Fig.~\ref{fig:msmt1}. The $d_m \times d_m$ patch corresponding to the qubit that needs to be read out using a faulty $T$ measurement is then left idling, until the pre-selection condition is satisfied and the qubit can be measured. While this patch is idling, a second $d_m \times d_m$ patch is used to perform the next $P_{\pi/8}$ rotation in step 2. With a high probability, the $d_m \times d_m$ patch left idling in step 1 will have been read out by the beginning of step 3, such that the freed-up space can be used to initialize a $d_m \times d_m$ patch for the next rotation. In rare cases, the $d_m \times d_m$ patch left idling in step 1 will have failed the pre-selection after, in this example, $2d_m$ code cycles, in which case it needs to keep idling, increasing the overall time cost of the distillation protocol. If this happens sufficiently rarely, the additional time cost is negligible.

The time required to perform a faulty $T$ measurement with pre-selection depends on the rejection rate. In the above example, if, e.g., the probability of failing the pre-selection is $50\%$ every two code cycles and $d_m=5$, then the probability of the $d_m \times d_m$ patch having failed the pre-selection after $2d_m$ code cycles is ${\sim}1\%$, implying that the overall time cost of \linebreak level-1 distillation protocols increases by ${\sim}1\%$ compared to the numbers presented in Tab.~\ref{tab:results}. This increase is even lower under the assumption of a rejection rate of ${\sim}4\%$. However, the numbers in Tab.~\ref{tab:results} do not take into account the increased space cost due to the additional $d_m \times d_m$ patches that are used to avoid the time-cost increase. If, as in Fig.~\ref{fig:msmt3}, one uses two $d_m \times d_m$ patches for each measurement region, then the space cost for, e.g., a $(\text{15-to-1})_{11,5,5}$ protocol increases by ${\sim}9\%$ compared to the numbers in Tab.~\ref{tab:results}. Note that, in two-level protocols, the increased space cost only affects the first level of distillation, but not the second, such that, e.g., for a $\text{(15-to-1)}_{11,5,5}^6 \times \text{(15-to-1)}_{25,11,11}$ protocol, the space cost increases by less than $2\%$. Furthermore, the error estimate in the main text does not take into account that, in cases where the $d_m \times d_m$ patches is left idling for many code cycles, the error rate of the corresponding $P_{\pi/8}$ rotations is increased. However, with pre-selection, the error rate of faulty $T$ measurements can be lower than the $p_{\rm phys}$ assumed in the main text, if single-qubit operations are significantly less noisy than two-qubit operations.

\textbf{Replacing faulty $T$ measurements with state injection.} After shrinking the $d \times d$ patch to a physical qubit, the preceding stabilizer measurement outcomes need to be decoded before a $T$ (or $T^\dagger$) gate and an $X$ measurement can be applied to the physical qubit. If the time required for this decoding operation is long compared to the decoherence time of idling physical qubits, the error rate of faulty $T$ measurements will be significantly increased. In this case, one should use state injection instead of faulty $T$ measurements, executing $P_{\pi/8}$ rotations via auto-corrected $\pi/8$ rotations~\cite{Litinski2019}, the same way $P_{\pi/8}$ rotations are executed in level-2 distillation protocols (see Fig.~\ref{fig:twolevel15to1layout}). This has a higher space cost compared to faulty $T$ measurements.

\textbf{What if faulty $T$ measurements have a higher error rate?} If one uses a simple faulty-$T$-measurement or state-injection protocol without post-selection of pre-selection, one might be worried that an increased $T$-measurement error rate could adversely affect the performance of the distillation protocols discussed in the main text. Specifically, we can consider the effect on the output error rate $p_{\rm out}$, if the $T$-measurement Pauli error rate is $10 p_{\rm phys}$ instead of $p_{\rm phys}$. We find that, for some of the distillation protocol shown in Tab.~\ref{tab:results}, an increased $T$-measurement error has only a small effect on the performance.

A collection of protocols is shown in Tab.~\ref{tab:higherr}. The main effect of an increased faulty-$T$-measurement error rate is an increase of the lowest achievable output error rate $p_{\rm out}$ for a given family of distillation protocols. For instance, the lowest achievable $p_{\rm out}$ for a one-level 15-to-1 protocol increases from $10.37p^3$ to $10370p^3$. For example, for $p_{\rm phys} = 10^{-4}$, a one-level 15-to-1 protocol can no longer produce magic states with $p_{\rm out} \approx 10^{-11}$ and, instead, a $(\text{15-to-1}) \times (\text{20-to-4})$ protocol needs to be used, significantly increasing the space-time cost to produce these states. On the other hand, the $\text{(15-to-1)}_{9,3,3}^4 \times \text{(20-to-4)}_{15,7,9}$ protocol produces states with a $p_{\rm out}$ that is far away from the lowest achievable $p_{\rm out}$ of a $\text{(15-to-1)} \times \text{(20-to-4)}$ protocol with $p_{\rm phys} = 10^{-4}$. Compared to Tab.~\ref{tab:results}, the output error rate increases from $p_{\rm out} = 2.4 \times 10^{-15}$ to $p_{\rm out} = 6.6 \times 10^{-15}$. Due to an increased failure probability, there is also a very small increase in space-time cost by $0.8\%$. Finally, $\text{(15-to-1)} \times \text{(15-to-1)}$ protocols can no longer produce magic states with an output error below $p_{\rm out} \approx 10^{-22}$.

For $p_{\rm phys} = 10^{-3}$, we find a similar trend. A one-level protocol can no longer be used to produce magic states with $p_{\rm out} \approx 10^{-8}$ and a more expensive two-level protocol needs to be used. For $p_{\rm out} = 2.5 \times 10^{-11}$, a $\text{(15-to-1)} \times \text{(20-to-4)}$ protocol needs to be replaced by a $\text{(15-to-1)} \times \text{(15-to-1)}$ protocol, increasing the space-time cost by $37.5\%$. Again, we find protocols which are only mildly affected by the increased error rate, such as the $\text{(15-to-1)}_{11,5,5}^6 \times \text{(15-to-1)}_{25,11,11}$ protocol, which previously produced magic states with an output error of $p_{\rm out} = 2.7 \times 10^{-12}$, but now produces magic states with an output error of $p_{\rm out} = 6.4 \times 10^{-12}$ with a space-time cost increase of $3.5\%$ due to the increased failure probability. The lowest achievable output error rate of $\text{(15-to-1)} \times \text{(15-to-1)}$ protocols increases to $p_{\rm out} \approx 10^{-14}$, making the production of magic states with error rates close to this limit very costly. For such states and states with a lower error rate, a three-level protocol might need to be used.

\end{document}